%% file: main.tex
\documentclass[reprint,superscriptaddress,nofootinbib,amsmath,amssymb,aps,pra,floatfix]{revtex4-1}
\usepackage{caption}
\usepackage{graphicx}
\usepackage{dcolumn}
\usepackage[inline]{enumitem}
\usepackage{hyperref}
\usepackage{pifont} 
\usepackage{cleveref}

\usepackage{bm}
\usepackage{amsmath}
\usepackage[font=small]{caption}
\usepackage[labelformat=empty, position=top]{subcaption}
\usepackage[export]{adjustbox}
\usepackage{blindtext}
\usepackage{amssymb}
\usepackage{newunicodechar}
\usepackage{nameref}

\begin{document}

\title{Adaptive Gated Graph Convolutional Network for Explainable Diagnosis of Alzheimer's Disease using EEG Data}

\author{Dominik Klepl}
    \affiliation{Centre for Computational Science and Mathematical Modelling, Coventry University, Coventry CV1 2JH, UK}

 \author{Fei He}
    \email{Corresponding authors: Fei He (fei.he@coventry.ac.uk) and Min Wu ()wumin@i2r.a-star.edu.sg}
    \affiliation{Centre for Computational Science and Mathematical Modelling, Coventry University, Coventry CV1 2JH, UK}

\author{Min Wu}
    \email{Corresponding authors: Fei He (fei.he@coventry.ac.uk) and Min Wu ()wumin@i2r.a-star.edu.sg}
    \affiliation{Institute for Infocomm Research (I$^{\text{2}}$R), Agency for Science, Technology and Research (A*STAR), Singapore)}

\author{Daniel J. Blackburn}
    \affiliation{Department of Neuroscience, University of Sheffield, Sheffield, S10 2HQ, UK}
  
\author{Ptolemaios G. Sarrigiannis}
    \affiliation{Department of Neurophysiology, Royal Devon and Exeter NHS Foundation Trust, Exeter, EX2 5DW, UK}
  

\begin{abstract}
Graph neural network (GNN) models are increasingly being used for the classification of electroencephalography (EEG) data. However, GNN-based diagnosis of neurological disorders, such as Alzheimer's disease (AD), remains a relatively unexplored area of research. Previous studies have relied on functional connectivity methods to infer brain graph structures and used simple GNN architectures for the diagnosis of AD. In this work, we propose a novel adaptive gated graph convolutional network (AGGCN) that can provide explainable predictions. AGGCN adaptively learns graph structures by combining convolution-based node feature enhancement with a correlation-based measure of power spectral density similarity. Furthermore, the gated graph convolution can dynamically weigh the contribution of various spatial scales. The proposed model achieves high accuracy in both eyes-closed and eyes-open conditions, indicating the stability of learned representations. Finally, we demonstrate that the proposed AGGCN model generates consistent explanations of its predictions that might be relevant for further study of AD-related alterations of brain networks.

\textit{Keywords:} Alzheimer's disease, graph neural network, classification, EEG
\end{abstract}

\maketitle

\section{Introduction}
The brain is a complex, densely connected system that operates across multiple spatial and temporal scales. Neurological diseases, such as Alzheimer's disease (AD), can alter the connectivity of the brain and thus disrupt brain function \cite{delbeuck2003disrupted, de2012disrupted, pievani2011functional, konig2005decreased}. AD is the most common cause of dementia and affects millions of patients worldwide \cite{mayeux2012epidemiology, rossini2020early}. Currently, the diagnosis of AD is typically made using a combination of cognitive and neurological assessments, as well as neuroimaging techniques, such as positron emission tomography (PET) or magnetic resonance imaging (MRI), which can be time-consuming and expensive \cite{babiloni2016synchronisationreview}. The development of rapid, economical, and explainable diagnosis methods is thus of importance \cite{rossini2020early}.

Electroencephalography (EEG) is an economical and non-invasive neuroimaging method that records the sum of electrical potentials generated by various brain areas. EEG is extensively used in the research of AD-related alterations in brain function and functional connectivity. Although EEG is not currently used in clinical settings for AD diagnosis, numerous studies have demonstrated the high effectiveness of an EEG-based diagnosis of AD \cite{klepl2022GNN, klepl2022bispectrum, klepl2021energylandscape, shan2022STGCN-AD, nobukawa2020classificationFC, fouladi2022contwaveletCNN}.

AD causes disruption of synaptic connections across multiple scales \cite{jeong2004eeg, pievani2011functional, klepl2021bispectrum} and can thus be viewed as a network disorder \cite{delbeuck2003disrupted}. The synaptic disconnection can be observed in EEG signals as alterations of synchronisation and functional connectivity (FC) \cite{pievani2011functional, babiloni2016synchronisationreview}. Furthermore, the slowing of EEG signals is a reliable characteristic of AD \cite{jeong2004eeg,ghorbanian2015slowing}, observed as a shift of spectral power towards low-frequency components. Graph-theoretic studies of AD also report reduced complexity, disruption of small-world properties, decreased integration, and increased segregation \cite{supekar2008smallworld, kabbara2018integrationsegregation, vecchio2021coherence-smallworld, stam2007correlation, klepl2021bispectrum, klepl2021energylandscape}. However, one of the challenges in EEG-based predictive models is the efficient utilisation of the information collected over multiple electrodes since there is information to be gained both at the electrode level, e.g. frequency spectrum, and the cross-electrode level, e.g. FC.

Machine learning-based approaches often require domain knowledge and rely on manual feature extraction. For example, Oltu et al. \cite{oltu2021PSD-coherence} calculate power spectrum density (PSD) and coherence across multiple EEG electrodes and then use descriptive statistics, such as sum and variance, as input features. Other feature-based methods use FC \cite{nobukawa2020classificationFC, song2018FC, yu2019fuzzyFC}. These methods first reconstruct the brain graph via measures of FC, such as phase lagging index \cite{nobukawa2020classificationFC}, generalised composite multiscale entropy vector \cite{song2018FC}, or phase synchronisation index \cite{yu2019fuzzyFC}. The features can then be extracted via statistics \cite{song2018FC} or graph-theoretic measures \cite{nobukawa2020classificationFC, yu2019fuzzyFC}.

In contrast, deep learning methods can extract features automatically from the input. However, utilising the information from multiple electrodes with classical deep learning methods is challenging. To overcome this issue, several studies have transformed EEG signals into images to make use of convolutional neural networks (CNN) \cite{ieracitano2019CNN, deepthi2020CNN, bi2019spectralimage, huggins2021deep, fouladi2022contwaveletCNN}, which are efficient in image classification. For instance,  Ieracitano et al. \cite{ieracitano2019CNN} compute the PSD across channels and compose them to form a channel by PSD image. Bi et al. \cite{bi2019spectralimage} use spectral topology images and leverage the colour channels of an image to represent three frequency bands. Finally, Huggins et al. \cite{huggins2021deep} create tiled images where each tile contains the continuous wavelet transform of an EEG electrode. Although these methods utilise multiple channels, the cross-electrode information is still omitted. A CNN trained on FC-based adjacency matrices has been proposed to address this limitation \cite{alves20FC-CNN}. However, CNN is not well suited for such input since the adjacency matrix is irregular and non-euclidean.

Graph neural network (GNN) is an extension of CNN to process graph-structured inputs. Multiple studies propose GNN-based architectures to process EEG. However, GNN methods for EEG-based diagnosis of AD are limited \cite{klepl2022GNN, shan2022STGCN-AD}. GNN-EEG implementations often include several steps: (1) input construction, i.e. graph structure and node features; (2) GNN encoder to learn node embeddings; and (3) aggregation of node embeddings to a graph embedding, which can be used in the final classification step.

There are various approaches to realise the graph construction in step (1). Node features are commonly defined as EEG time-series signal \cite{shan2022STGCN-AD,li2021multidomain, asadzadeh2022bayes, li2022bimodal}, or a statistical summary of the signal in the time domain \cite{raeisi2022neonatal-epilepsy, tang2021selfsupervised}, the frequency domain \cite{klepl2022GNN, sun2022transformer}, or the differential entropy \cite{li2021multidomain, yin2021LSTMfusion, chen2021bayesianGNN, jia2020graphsleepnet, zeng2022siam, sun2022transformer}. Based on network neuroscience literature, many approaches define the brain graph using FC measures \cite{klepl2022GNN, shan2022STGCN-AD, liu2022MST-GNN, chang2021MMN-schizophrenia, raeisi2022neonatal-epilepsy, tang2021selfsupervised, li2021multidomain, li2022bimodal}. The graph structure can also be based on the distance between EEG electrodes \cite{tang2021selfsupervised, yin2021LSTMfusion, chen2021bayesianGNN}. However, such an approach largely ignores brain connectivity information. Alternatively, the brain graph can be automatically learned by the model, either as a learnable mask shared across samples \cite{li2019GNN-motor-movement,li2021multidomain,sun2022transformer} or by pairwise node feature distance minimisation regularised by an additional graph loss function \cite{zhong2020GNN-regularised-emotion,jia2020graphsleepnet,zeng2022siam}. While such approaches are flexible and should converge to an optimal graph structure with respect to a given learning task, the learned brain graph might not be representative of the underlying brain connectivity, i.e. such a graph structure might overestimate the strength of the task-relevant edges compared to the underlying connectivity. In this work, we propose an adaptive graph learning mechanism based on node feature enhancement via CNN and subsequent graph construction. This is achieved by using a correlation similarity measure of power spectral density and sparsified via k-nearest neighbour (KNN) edge selection. Thus, it combines the strength of the FC-based and automated graph learning methods. Such a combination overcomes the limitations of fully learnable graphs described above since the correlation computation is ultimately detached from the classification task. However, it should be noted that the adaptively learned graph structure reflects brain region similarity rather than a functional relationship assumed by classical FC measures.

The design of GNN encoders in step (2) for EEG applications has been mainly limited to simple architectures, such as the Chebyshev graph convolution (ChebConv) \cite{chang2021MMN-schizophrenia,asadzadeh2022bayes,tang2021selfsupervised,yin2021LSTMfusion,chen2021bayesianGNN,jia2020graphsleepnet,li2022bimodal}, and simple graph convolution (GCN) \cite{klepl2022GNN,li2019GNN-motor-movement,zhdanov2022schizophrenia,li2021multidomain,sun2022transformer,zhong2020GNN-regularised-emotion}. However, we hypothesise that such node embedding updating mechanisms are not optimal for EEG tasks. These graph convolutions update node embeddings by summing the initial embedding and the aggregated messages from the neighbouring nodes. Such updating implies that information from different scales contributes equally to the final node embeddings, hence graph embeddings as well. While brain disruptions caused by AD occur across multiple spatial scales, their predictive power is likely different. Therefore, a gating mechanism is crucial for filtering and weighting the information collected across different scales. We propose to adopt the gated graph convolution \cite{li2016GRGCN} to address this issue.

Finally, we implement the aggregation of node embeddings in step (3) by adopting the adaptive structure-aware pooling (ASAP) node pooling mechanism \cite{ranjan2020asap} to first learn the most important clusters of nodes, which are in turn concatenated to form the graph embedding. This is in contrast to the previous approaches that do not use any node pooling and form graph embeddings via simple element-wise readout layers \cite{klepl2022GNN, xu2022dagam,raeisi2022neonatal-epilepsy,li2021multidomain,zhdanov2022schizophrenia,liu2022MST-GNN,zhong2020GNN-regularised-emotion} or concatenating all nodes of the graph \cite{shan2022STGCN-AD, chang2021MMN-schizophrenia}. Other node pooling approaches were tested for EEG applications \cite{demir2021ERP-comparison,xu2022dagam}. In contrast to ASAP pooling, these approaches pool the graph by selecting a specified number of nodes without considering their local context within the graph. Therefore, important information might be lost due to such node pooling.

\begin{figure*}[t]
    \centering
    \includegraphics[width = 0.98\linewidth]{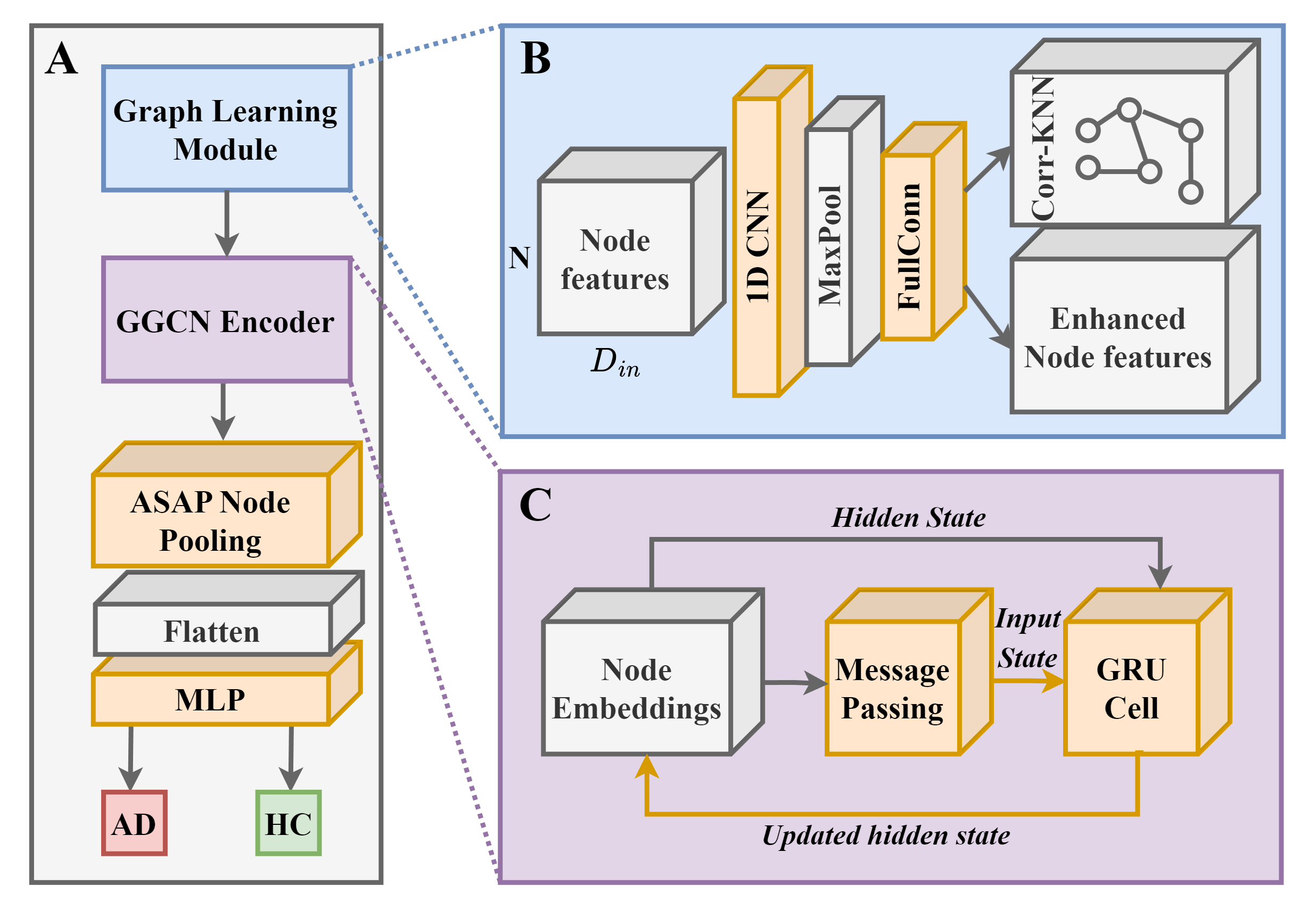}
    \caption{The architecture of the proposed adaptive gated graph convolutional network. A) The proposed model consists of a graph learning module, GGCN encoder, ASAP node pooling module, and a three-layer multilayer perception (MLP) outputting the predicted probabilities. B) Graph learning module takes a $N \times D_{in}$ node feature matrix as input. Node features are defined as power spectral density from 1 to 45 Hz ($D_{in}=45$) computed for all $N$ EEG electrodes ($N=23$). Then, a 1D CNN enhances them. The brain graph structure is then constructed as a correlation graph between the outputs from the 1D CNN and made sparse by a k-nearest-neighbour edge selection (Corr-KNN). C) The enhanced node features and the learned graph structure are then passed to a gated graph convolutional neural network (GGCN) encoder. GGCN applies message passing and gated recurrent unit (GRU) recursively over $R$ iterations.}
    \label{fig:overview}
\end{figure*}

In this paper, we propose a novel GNN model for explainable AD classification, which can adaptively enhance node features and dynamically construct brain graph structures as shown in Fig. \ref{fig:overview}. The learned brain graphs can then be used for the interpretation of predictions. Moreover, a clustering-based node pooling mechanism is adopted to coarsen the brain graph, thus localising the brain regions that contribute to the predictions. Finally, we conduct extensive ablation and parameter sensitivity experiments to elucidate the importance of the individual blocks within the proposed model architecture.

\section{Data}
EEG recordings were collected from 20 AD patients and 20 healthy control participants (HC) younger than 70 years. A detailed description of the experimental design and confirmation of the diagnosis is provided in \cite{blackburn2018synchronisation}. All the AD participants were recruited from the Sheffield Teaching Hospital memory clinic. AD participants were diagnosed between one month and two years before data collection. All of them were in the mild to moderate stage of the disease at the time of recording, with an average Mini Mental State Examination (MMSE) score of 20.1 (sd = 4). High-resolution structural magnetic resonance imaging (MRI) scans of all patients were acquired to eliminate alternative causes of dementia. 
Age and gender-matched HC participants with normal neuropsychological tests and structural MRI scans were recruited.
This study was approved by the Yorkshire and The Humber (Leeds West) Research Ethics Committee (reference number 14/YH/1070). All participants gave their informed written consent.

EEG data were acquired using an XLTEK 128-channel headbox, Ag/AgCL electrodes with a sampling frequency of 2 kHz using a modified 10-10 overlapping a 10-20 international electrode placement system with a referential montage with a linked earlobe reference. The recordings lasted 30 minutes, during which the participants were instructed to rest and not think about anything specific. In case the participants showed signs of drowsiness, they were prompted. Within each recording were five-minute-long epochs during which the participants had their eyes closed, alternating with an equal duration of eyes-open epochs.

All the recordings were reviewed by an experienced neurophysiologist on the XLTEK review station with time-locked video recordings (Optima Medical LTD). For each participant, three 12-second-long artefact-free epochs were isolated. Finally, the following 23 bipolar channels were created: F8–F4, F7–F3, F4–C4, F3–C3, F4–FZ, FZ–CZ, F3–FZ, T4–C4, T3–C3, C4–CZ, C3–CZ, CZ–PZ, C4–P4, C3–P3, T4–T6, T3–T5, P4–PZ, P3–PZ, T6–O2, T5–O1, P4–O2, P3–O1 and O1–O2 \cite{blackburn2018synchronisation}. Bipolar montage was selected to limit the volume conduction effects to a certain extent.

As a neurophysiologist confirmed the EEG signal to be artefact-free, we did not further clean the signals. The signals are filtered using a band-pass Butterworth filter to a range of 0.5 Hz and 45 Hz and down-sampled to 200 Hz. Finally, 1-second long windows with 50\% overlap are created to increase the sample size.

\section{Methods}
The proposed adaptive gated graph convolutional network (AGGCN) model consists of three blocks: a graph learning module, a GNN encoder and a classifier. The graph learning module receives a node feature matrix as input, enhances it using a 1D-CNN and learns the brain graph structure. The GNN encoder then uses the output of the graph learning module as input, i.e. a featured, weighted, undirected graph. The encoder generates a graph embedding used by the classifier to output the predicted probabilities.

\subsection{Node feature and graph learning}
The node features are defined as power spectral density computed from 1-second-long EEG signals with 1 Hz increments from 1 to 45 Hz. Hence, the input is a node feature matrix $\mathbf{X} \in \mathbb{R}^{N \times D_{in}}, D_{in}=45$. 

The input is then passed to a convolutional neural network (CNN) with batch normalisation, $L_{CNN}$ 1D convolutional layers and a maximum pooling with kernel size 2 and step size 2. The output is flattened and fed to a fully connected layer with hidden size $h_{CNN}$ and batch normalisation. This neural network outputs a matrix of enhanced node features $\mathbf{X'} \in \mathbb{R}^{N \times D_{h_{CNN}}}$.

A graph structure is then inferred from the enhanced node features by computing the absolute value of Pearson's correlation for each pair of nodes. Thus, a unique graph structure is learned for each input sample and is defined by an adjacency matrix $\mathbf{A} \in \mathbb{R}^{N \times N}$ with $N=23$ being the number of EEG channels. In order to produce sparse graphs, the k-nearest-neighbours algorithm is utilised. This means that the $k$ strongest edges are preserved for each node. 

This proposed graph learning module has multiple hyperparameters that control its architecture. Namely, these are the number of convolutional layers $L_{CNN}$, the kernel size (which is equal to the step size), the number of filters, the hidden size $h_{CNN}$, the dropout rate $drop_{CNN}$ and the $k_{KNN}$ parameter that controls the graph sparsity.

\subsection{Graph neural network encoder and classifier}
A graph convolution extends the classical convolution from the Euclidean domain to the graph domain. The input graph is given by $G=(\mathbf{N}, \mathbf{A}, \mathbf{X'})$ where $N$ is the set of nodes, $\mathbf{A}$ is the learned graph, and $\mathbf{X'}$ is the enhanced node feature matrix. A simple graph convolution is defined by the message-passing mechanism wherein the node embedding of node $i$ is learned by aggregating information from its 1-hop neighbourhood, i.e. nodes connected with an edge, as follows:
\begin{equation}
    \mathbf{x^{l+1}_{i}} = \mathbf{x^{l}_i} +\mathbf{\Theta}\sum_{j \in N(i)} e_{ij} \mathbf{x^{l}_j},
\end{equation}
where $\mathbf{x^{l}_{i}}$ are the node features of node $i$ at the $l$\textsuperscript{th} layer, $\mathbf{x^{0}_{i}}$ is the $i$\textsuperscript{th} row of the input node feature matrix $\mathbf{X}$, and $\mathbf{\Theta}$ is a learnable linear transformation. $N(i)$ and $e_{ij}$ are the neighbourhood of node $i$ and the edge weight connecting nodes $i$ and $j$ given by the adjacency matrix $\mathbf{A}$, respectively. Stacking $L$ graph convolutional layers then means aggregating information iteratively from 1-hop to $L$-hop neighbourhoods, thus gradually going from local to global information about the graph.

Note that the aggregated message is added to the initial node embedding $\mathbf{x^l_i}$. Thus, the entire information collected from each L-hop neighbourhood is always fully integrated into the node embedding. However, information might be distributed unequally across spatial scales in brain graphs. The gated graph convolution (GGCN) \cite{li2016GRGCN} addresses this problem by introducing a mechanism to decide what information should be retained at each scale selectively:
\begin{align} 
\mathbf{m_i^{(r+1)}} &= \sum_{j \in N(i)} e_{ji} \cdot \mathbf{\Theta^{r+1}} \cdot \mathbf{x_j^{(r)}}, \\
\mathbf{x_i^{(r+1)}} &= \textrm{GRU} (\mathbf{m_i^{(r+1)}},\mathbf{x_i^{(r)}}),
\label{eq: GGCN}
\end{align}
where $\mathbf{m_i}$ are the aggregated messages, $\sum$ is the aggregation function, $\mathbf{\Theta^{r}}$ is a learnable matrix for iteration $r$, which maps the node features from shape $[1, D_{h_{CNN}}]$ to $[1, D_{h_{GNN}}]$, and $GRU$ is the gated recurrent unit \cite{cho2014GRU}. Briefly, a GRU is a recurrent neural network layer with update, reset, and input gates that allow the network to recursively update or forget information about the input. The node embeddings are learned recursively up to $R$ iterations with a shared $GRU$ gate, which is equivalent to stacking $R$ GCN layers.

The node embeddings are then passed through an activation function and a batch normalisation layer. Finally, the node embeddings are passed to the node pooling module. The hyperparameters of the proposed encoder are the number of iterations $R$, the hidden size $h_{GNN}$, the activation function, the aggregation function and the dropout rate $drop_{GNN}$ applied after the encoder.

\subsubsection{Node pooling}
After learning the node embeddings, the model learns a coarsened graph using the ASAP pooling mechanism \cite{ranjan2020asap}. This pooling first learns $N$ clusters, each centred at one node, also named ego-graphs. The membership of node $j$ in the ego-cluster centred at node $i$ is given by the $\mathbf{S_{ij}}$ matrix. Note that this is a soft-cluster assignment matrix; thus, each node can belong to multiple clusters with varying membership strengths. The clusters are learned as follows:
\begin{align}
S_{ij} &= a_{ij}, \\
a_{ij} &= \mathrm{softmax} \left( \mathbf{\theta}^{\mathrm{T}} \sigma\left(\mathbf{\Theta} \mathbf{x^m_i} \Vert \mathbf{x_j}\right) \right), \label{eq: attention} \\
\mathbf{x^m_i} &= \max_{j \in N(i)} \mathbf{x_j},
\end{align}
where $a_{ij}$ is the attention score and the membership strength, $\mathbf{\theta}$ and $\mathbf{\Theta}$ are learnable vector and matrix, respectively. $\sigma$ is the LeakyReLU activation function, and $\mathbf{x^m_i}$ is the master query representing the initial cluster embedding. The attention scores are also subject to a dropout probability $drop_{pool}$. The final cluster embedding is then calculated as an attention-weighted sum, which is additionally weighted by the cluster score $\phi_i$:
\begin{equation}
\mathbf{x^c_i} = \phi_i \sum_{j \in N(i)} a_{ij}\mathbf{x_j},
\label{eq: attention weighted sum}
\end{equation}
where the cluster score $\phi_i$ is computed by the local extremum graph convolution \cite{ranjan2020asap}:
\begin{equation}
    \phi_i = \mathbf{\Theta_1} \cdot \mathbf{x_i} + \sum_{j \in N(i)} e_{ji} \cdot \left( \mathbf{\Theta_{2}}\mathbf{x_i} - \mathbf{\Theta_{3}}\mathbf{x_j} \right),
\end{equation}
which is designed to measure the relative importance of each cluster. 

The cluster embedding $\mathbf{x^c_i}$ is then used to select the top $k$ scoring clusters, which will be included in the coarsened graph:
\begin{align}
\bar{i} &= Top_k(\mathbf{X^c}), k \in [1,2,...N], & \bar{\mathbf{S}} &= \mathbf{S}(:,\bar{i}) \\
\mathbf{A^{p}} &=\mathbf{\bar{S}^\mathrm{T}}\cdot \mathbf{A}\cdot\mathbf{\bar{S}}, & \mathbf{X^{\mathrm{p}}} &= \mathbf{X^c}(:,\bar{i})
\end{align}
where $Top_k$ is a function that returns the indices of clusters $\bar{i}$. $\mathbf{\bar{S}}$ and $\mathbf{X^{\mathrm{p}}}$ are the pruned soft-cluster assignment matrix and the pruned cluster embedding matrix, respectively, and $\mathbf{A^{p}}$ is the adjacency matrix of the coarsened graph.

The graph pooling module has the following hyperparameters: the size of the pooled graph $k_{pool}$, the dropout rate $drop_{pool}$ and the negative slope of the LeakyReLU activation.

\subsubsection{Multilayer perceptron classifier}
The cluster embedding matrix $\mathbf{X^p}$ of the coarsened graph returned by the node pooling module is flattened and fed to a multilayer perceptron (MLP) classifier. Specifically, a $L_{MLP}$-layer MLP with hidden size $h_{MLP}$ is utilised with a block of batch normalisation, activation function, and dropout layers utilised between the fully connected layers. The final layer outputs a two-dimensional vector of log probabilities for each class.

The classifier has the following hyperparameters: the number of layers $L_{MLP}$, hidden size $h_{MLP}$, activation function and dropout rate $drop_{MLP}$.

\begin{table*}[ht]
\centering
\caption{Performance of the proposed AGGCN in eyes closed (EC), eyes open (EO) and combined (EC+EO) conditions.}
\begin{tabular}{cccccc}
  \hline
Condition & Accuracy & AUC & Sensitivity & Specificity & F1 \\ 
  \hline
EC & 89.1  $\pm$  1.4 & 0.895  $\pm$  0.016 & 92.95  $\pm$  2.59 & 85.16  $\pm$  2.45 & 89.7  $\pm$  1.4 \\ 
  EO & 85.56  $\pm$  0.96 & 0.834  $\pm$  0.015 & 90.88  $\pm$  2.01 & 79.98  $\pm$  1.47 & 86.55  $\pm$  0.98 \\ 
  EC+EO & 81.79  $\pm$  1.26 & 0.82  $\pm$  0.016 & 84.27  $\pm$  2.19 & 79.22  $\pm$  2.05 & 82.46  $\pm$  1.27 \\ 
   \hline
\end{tabular}
\label{table:performance-conditions}
\end{table*}

\subsection{Model implementation and evaluation}
The proposed AGGCN model was implemented using PyTorch 1.10 \cite{pytorch}, and PyTorch Geometric 2.0.2 \cite{pytorch-geometric} and trained on a laptop with Intel i7 CPU, 16 GB RAM and an NVIDIA RTX 2070 GPU.

The model is trained by minimising the cross-entropy loss. The model performance is evaluated using repeated (30 times) 10-fold stratified group cross-validation (one group = subject identifier) and trained on the dataset collected during the eyes-closed condition. Since all participants have multiple samples, keeping all the samples from the same participant within the same fold is crucial to prevent information leakage. In order to prevent overfitting, another fold is utilised for validation to implement early stopping and is used to optimise hyperparameters. Thus, in each iteration of the cross-validation, one fold is used as validation, one fold as testing, and the remaining eight folds form the training set.

A stochastic gradient descent (SGD) optimiser and an exponential learning rate scheduler are used to train the model with a batch size of 128 for 200 epochs. If validation loss does not decrease for 15 epochs, the training is stopped early. Additionally, zero-mean Gaussian noise with standard deviation $\sigma$ is added to the input during training with probability $p_{noise}$ to improve the generalisability of the model. Eventually, the best model was identified using the average cross-validated F1 score measured on the validation folds. The selected model was then retrained and tested on the dataset obtained during the eyes-open condition and the combined dataset from both conditions. The final results are then reported using the test folds only. The stability of the performance is assessed by computing the standard deviation of the samples collected over the 30-times repeated cross-validation.

Note that the hyperparameters of the proposed model are optimised using Bayesian optimisation. Ten warm-up random iterations were used to initialise the optimisation, followed by 200 optimisation iterations. The optimisation is evaluated only on the validation sets to prevent overfitting. Moreover, we carry out parameter-sensitivity experiments to verify the influence of a few key hyperparameters of the proposed model architecture. Specifically, these are the number of iterations of the GGCN encoder, the size of the pooled graphs, the sparsity of the learned graph and the choice of aggregation function of the GGCN encoder. Due to the computational cost of running these experiments, we reduce the number of repeats of the cross-validation from 30 to 5. The hyperparameters of the model are reported in our supplementary materials.

\section{Results and Discussion}
In this section, we report the experimental results of our AGGCN model. As illustrated in Table \ref{table:performance-conditions}, our AGGCN has shown robust performance across all the conditions. Note that the best performance was achieved during the EC condition. This is likely because with eyes closed, the ocular artefacts are minimised; thus, the underlying dynamics are easier to detect. The performance remains high even in the EO condition, suggesting that the proposed model can detect underlying patterns in both EC and EO conditions. However, the performance decreases significantly on the EC+EO combined dataset. We hypothesise that the patterns learned under the EC and EO conditions share relatively little information; thus, the EC+EO model performs significantly worse. We explore this further in section \ref{section: explainability}.

In addition, the hyperparameter values of the optimised model are reported in Table III in Supplementary Materials.

\subsection{Comparison with the baselines}

The proposed model was compared to seven baseline models proposed in the literature across the three conditions. The first baseline is the best-performing model from our previous work \cite{klepl2022GNN}. It is a GNN with two spatial graph convolutional layers, maximum readout and brain graph defined using the amplitude-envelope-correlation (AEC-GNN). The second baseline model is the spatio-temporal GNN (STGCN) that uses temporal convolutions and ChebConv layers and defines the brain graphs using wavelet coherence \cite{shan2022STGCN-AD}. Then, two CNN-based models, PSD-CNN \cite{ieracitano2019CNN} and Wavelet-CNN \cite{huggins2021deep}, trained on PSD and wavelet transform, respectively, were used. Next, two traditional machine learning approaches were utilised: support vector machine trained on node degree computed from phase lag index graph (NS-SVM) \cite{nobukawa2020classificationFC}, and a logistic regression trained on vectorised adjacency matrices obtained from coherence graphs across seven frequency bands (AM-SVM) \cite{musaeus_oscillatory_2019}. Finally, we use an MLP model where the input is a flattened PSD node feature matrix \cite{klepl2022GNN} without using graph-domain information.

Table \ref{table:baseline-performance} shows the f1 scores of various models across different conditions. Note that all seven models were evaluated under the same setting (e.g. the same 1-second EEG window samples). We can observe that our proposed AGGCN outperforms the baselines across all conditions.
Moreover, STGCN was originally evaluated using a cross-validation setup, which mixed samples from the same subject in their original paper.\cite{shan2022STGCN-AD}. It is expected that its performance drops significantly when evaluated using stratified group cross-validation in our experiments. 

\begin{table*}[ht]
\centering
\caption{The F1 score and the number of trainable parameters of the baseline models and the proposed method across conditions. The best-performing model is highlighted in bold.}
\begin{tabular}{ccccc}
  \hline
Model & EC & EO & EC+EO & No. of parameters \\ 
  \hline
AEC \cite{klepl2022GNN} & 81.61  $\pm$  3.16 & 77.91  $\pm$  1.1 & 76.74  $\pm$  1.87 & 445,204 \\ 
  MLP \cite{klepl2022GNN} & 82.01  $\pm$  4.39 & 76.51  $\pm$  3.34 & 77.47  $\pm$  4.26 & 54,628,354 \\ 
  PSD-CNN \cite{ieracitano2019CNN} & 88.15  $\pm$  0.77 & 80.89  $\pm$  1.45 & 79.51  $\pm$  1.74 & 3,420,432 \\ 
  STGNN \cite{shan2022STGCN-AD} & 46.71  $\pm$  8.58 & 44.34  $\pm$  7.33 & 38.25  $\pm$  17.16 & 662,754 \\ 
  Wavelet-CNN \cite{huggins2021deep} & 51.35  $\pm$  5.61 & 57.52  $\pm$  8.02 & 59.27  $\pm$  6.44 & 46,755,208 \\ 
  AM-SVM \cite{musaeus_oscillatory_2019} & 86.3 $\pm$ 1.5 & 83.8 $\pm$ 1.3 & 80.31 $\pm$ 1.3 & \ding{56} \\ 
  NS-SVM \cite{nobukawa2020classificationFC} & 55.93 $\pm$ 3.04 & 50.32 $\pm$ 3.36 & 52.9 $\pm$ 2.08 & \ding{56} \\ 
  Proposed & \textbf{89.7  $\pm$  1.4} & \textbf{86.55  $\pm$  0.98} & \textbf{82.46  $\pm$  1.27} & 2,208,861 \\ 
   \hline
\end{tabular}
\label{table:baseline-performance}
\end{table*}

\subsection{Model ablation study}

We perform ablation experiments to determine the contribution of each module of the proposed model. The following seven ablated variants of the proposed model were tested in our experiments.

\begin{itemize}
    \item \textbf{A}: no node pooling;
    \item \textbf{B}: graph learning replaced with a fully connected graph;
    \item \textbf{C}: GGCN replaced with a $R$\textsuperscript{th}-order ChebConv ($R=4$);
    \item \textbf{D}: variants A and B combined;
    \item \textbf{E}: variants A and C combined;
    \item \textbf{F}: variants B and C combined;
    \item \textbf{G}: variants A, B and C combined.
\end{itemize}

\begin{figure}[ht]
    \centering
    \includegraphics[width = 0.98\linewidth]{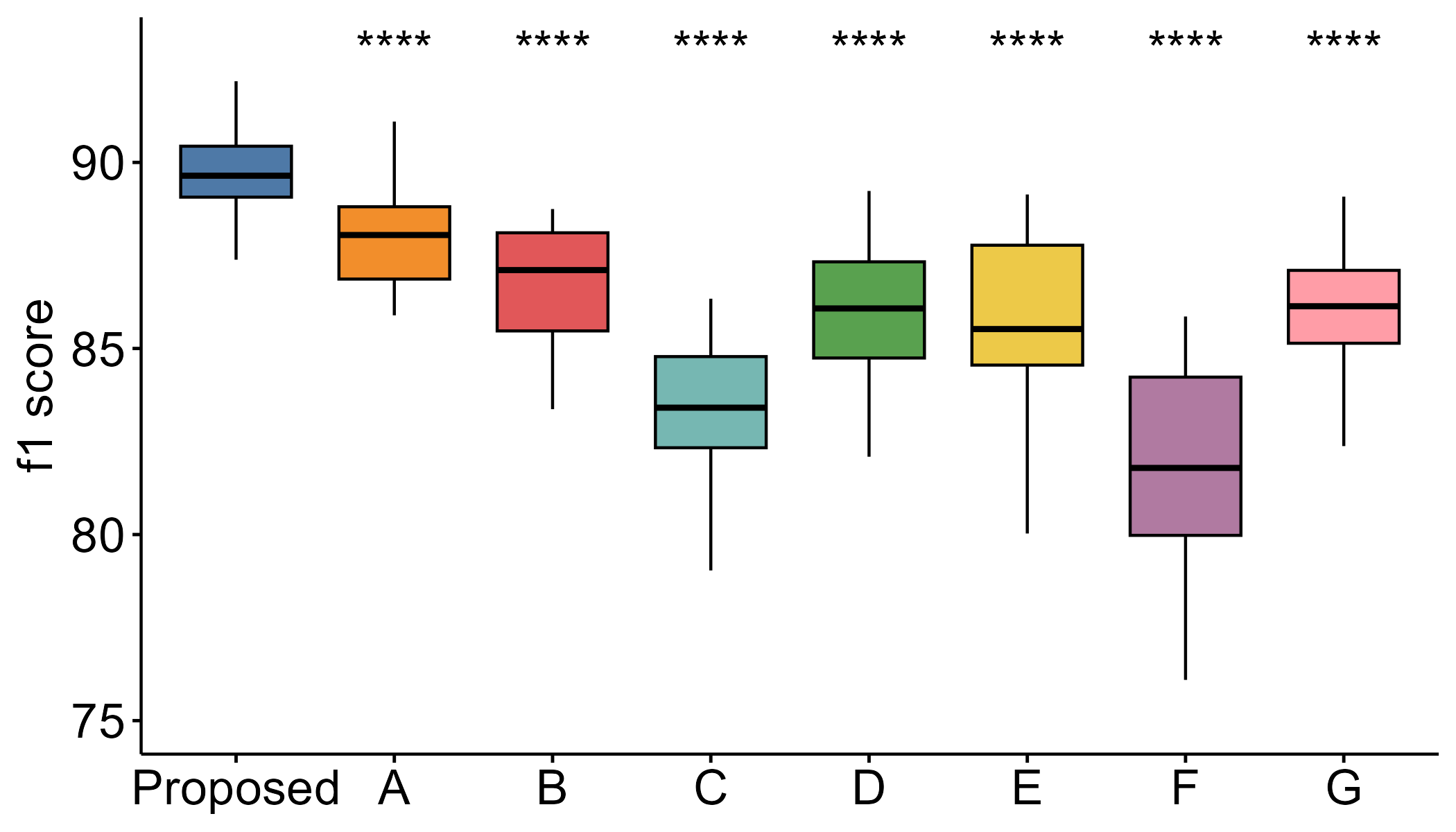}
    \caption{F1 scores of model variants. The asterisks report the p-value of a nonparametric Mann-Whitney U test measuring the difference between AGGCN and the ablated variants.}
    \label{fig:ablation}
\end{figure}

The ablation results in Fig. \ref{fig:ablation} reveal that each of the proposed modules contributes significantly to the high performance of the proposed architecture. For variant A, we can observe that the contribution of the node pooling module is significant, albeit relatively small. However, this module reduces the number of parameters of the model and helps to produce explainable predictions (Fig.  \ref{fig:cluster_probability} and Fig. \ref{fig:cluster_attention}). Without the node pooling, the final MLP classifier would have $N \times h_{GNN} \times h_{MLP}$ parameters ($N=23$), but node pooling reduces it to $k_{pool} \times h_{GNN} \times h_{MLP}$ ($k_{pool}=3$). For variant B, it is not surprising that its performance decreases significantly as the graph learning module is replaced with a fully connected graph. Thus, it cannot leverage graph-domain information except in the node pooling module. 

Next, we demonstrate that the GGCN encoder improves performance significantly compared to a ChebConv encoder according to variant C. A ChebConv layer is similar to a GGCN in its iterative nature, i.e. ChebConv iteratively updates node embeddings by approximating the eigendecomposition of graph Laplacian. However, ChebConv does not have any gating mechanism, which means that information across scales contributes to the final embedding equally. Since all of the major modules of the proposed are shown to contribute to the final performance significantly, it is unsurprising that the rest of the ablated models with more than one of these modules perform significantly worse as well (Variants D-G in Fig. \ref{fig:ablation}). Note that some of the ablated models maintain a relatively low variance of performance. We speculate this is because the ablated models can still learn robust embeddings, but some of the information within the data remains inaccessible, which would be enabled by the removed module.

The parameter sensitivity experiments also support the optimal values of crucial hyperparameters of the proposed model (Supplementary Materials, Figs 10-13). It is worth noting that the proposed architecture allows training relatively deep models (using up to eleven GGCN iterations) with only a minor performance decrease (Fig. 10). We can also observe that although the optimal values of these hyperparameters result in the best performance, the performance doesn't change much with adjacent values near the optima. This demonstrates that although the proposed model requires a relatively large number of hyperparameters to be determined, its performance remains robust with sub-optimal values, thus suggesting generalisability potential.

\subsection{Explainability of AGGCN}
\label{section: explainability}
The proposed model generates plausible and consistent explanations for its predictions. We generate multiple types of prediction explanations. Specifically, these are derived from the following: (1) graph learning, (2) node embedding and GGCN encoder, (3) node pooling, and (4) feature masking. Except for type (4), these explanations could be obtained for individual samples. However, we visualise the diagnosis-averaged explanations to explore the patterns learned by the proposed model.

\subsubsection{Graph learning}
\begin{figure}[ht]
    \centering
    \includegraphics[width = 1.05\linewidth]{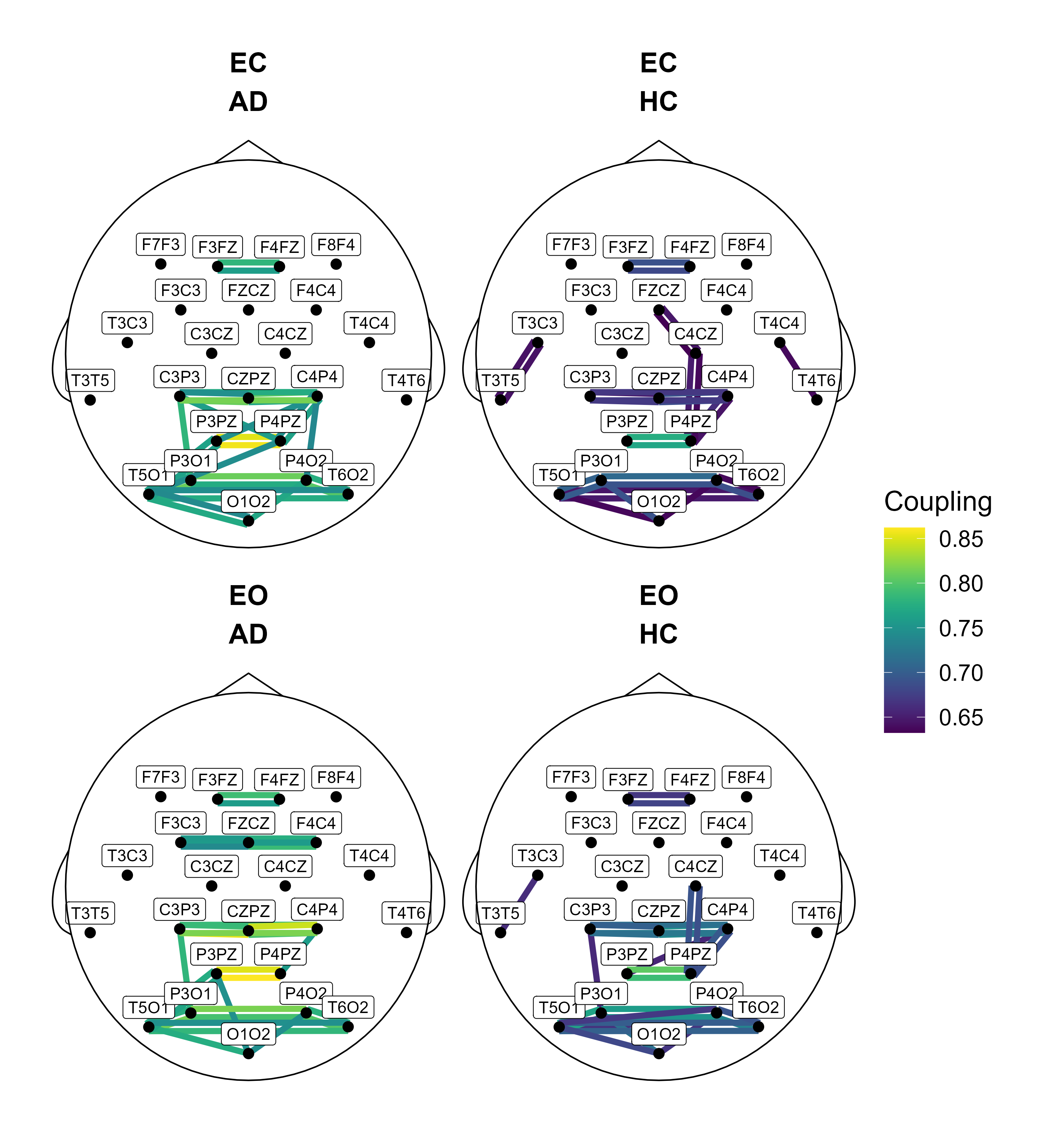}
    \caption{Top 30 strongest edges of the AGGCN-learned graphs of AD and HC cases in EC and EO conditions (average of all samples).}
    \label{fig:learned_graph}
\end{figure}

\begin{figure}[ht]
    \centering
    \hspace*{\fill}
    \includegraphics[width = 1.05\linewidth]{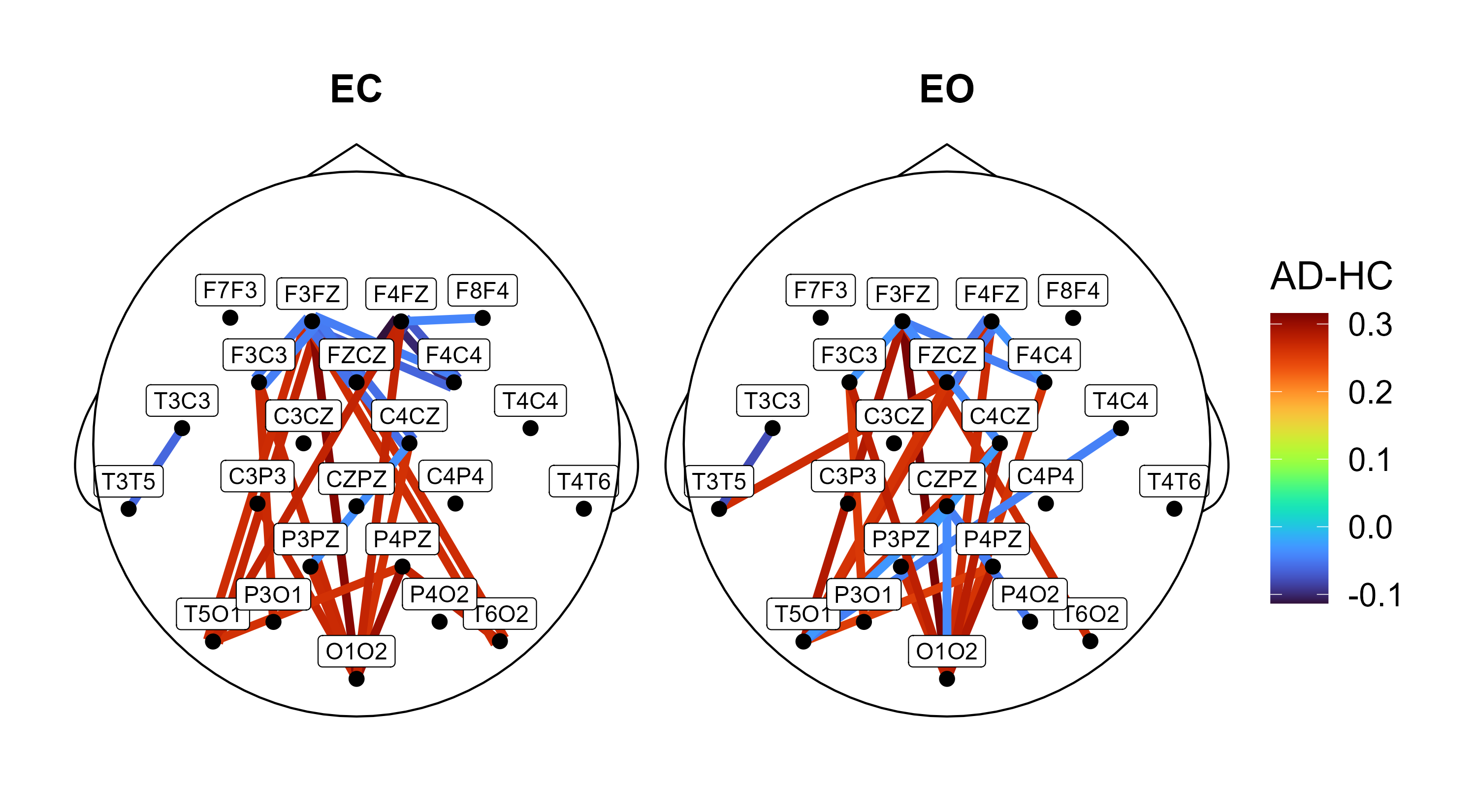}
    \caption{The differences between AGGCN-learned graphs for AD and HC cases in EC and EO conditions show the AD-related connectivity disruption. The average of all samples, the top 30 strongest edges were preserved. Values above zero indicate AD increase, while values below zero indicate AD decrease.}
    \label{fig:learned_graph_difference}
\end{figure}

The graph learning module learns a clear difference between the AD and HC cases, as shown in Fig. \ref{fig:learned_graph} (alternatively Fig. 14). The learned brain graphs show that AD cases have increased connectivity overall, while HC graphs seem more sparse with few densely connected regions. A well-defined cluster of densely connected nodes is present in both groups within the centro-parietal and occipital regions and a few strong edges in the frontal and temporal regions. The locations of the strongest edges are consistent across conditions. Fig. \ref{fig:learned_graph_difference} then shows the top 30 edges, where the largest increase/decrease in coupling was observed in AD. AD seems to have increased coupling strength in long-distance edges, particularly between frontal and parietal/occipital regions. These increases are quite consistent between conditions. In contrast, AD cases have decreased coupling strength, mostly in local connections in the frontal (EC) and frontal and centro-parietal (EO) regions.

Additionally, we statistically compared the learned graph structures to determine differences between AD and HC cases across EC and EO conditions. The results of this analysis are reported in the supplementary materials (Fig 15).

\subsubsection{Node embeddings and GGCN}
\begin{figure}[ht]
    \centering
    \includegraphics[width = 1.05\linewidth]{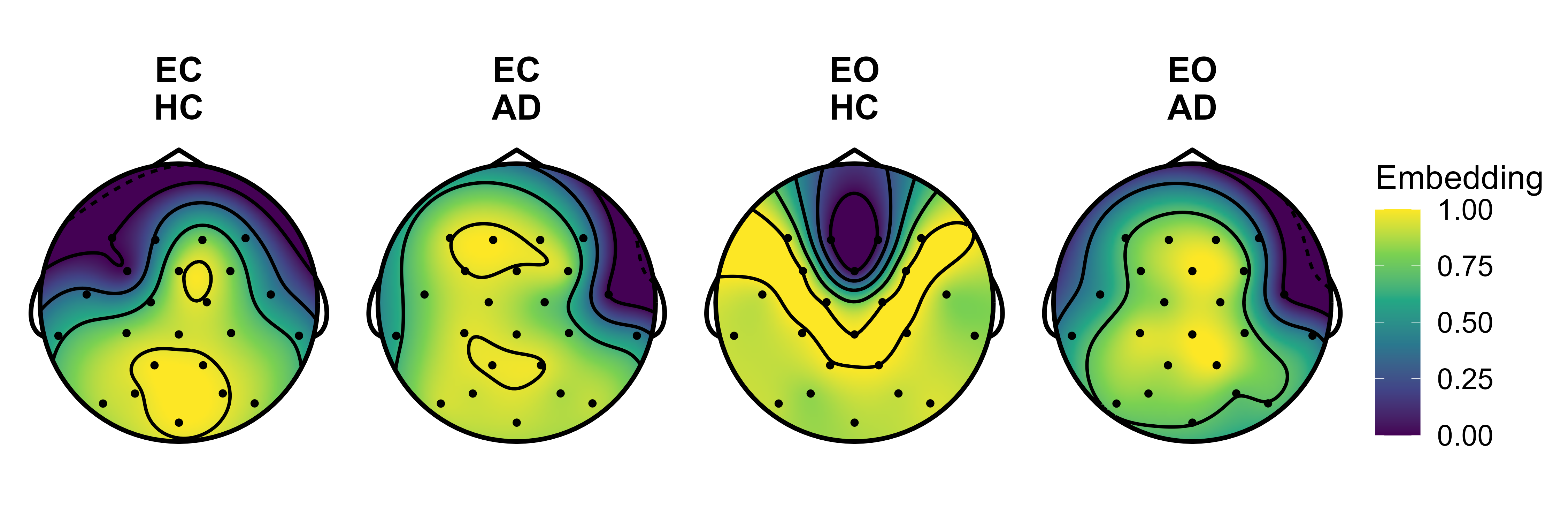}
    \caption{Averaged node embeddings across nodes expressed via the first component of PCA for AD and HC cases across EC and EO conditions. Note that embedding value does not suggest increased or decreased activity within a given area but rather the similarity of nodes.}
    \label{fig:node_embedding}
\end{figure}

Another prediction explanation can be derived from the node embeddings obtained by the GGCN (Fig \ref{fig:node_embedding}). In particular, we visualise the node embeddings obtained after four iterations of GGCN and compress them to 1D representation using principal component analysis (PCA) and extracting the first principal component. PCA is fitted for each condition separately. The node embeddings do not express a change in activity but rather a node similarity. Generally, the node embedding explanations show two large regions of similar embeddings. In EC, these are frontotemporal and centro-parietal regions, and right frontotemporal and the rest of the regions for HC and AD, respectively. The HC similarity region in the EO condition is reduced from frontotemporal to only the frontal region. In contrast, the AD similarity region expands from the right frontotemporal region to the left side. This further highlights the differences in learned patterns under the EC and EO conditions, thus explaining the reduced performance in the combined EC+EO condition.

\begin{figure}[ht]
    \centering
    \includegraphics[width = 0.98\linewidth]{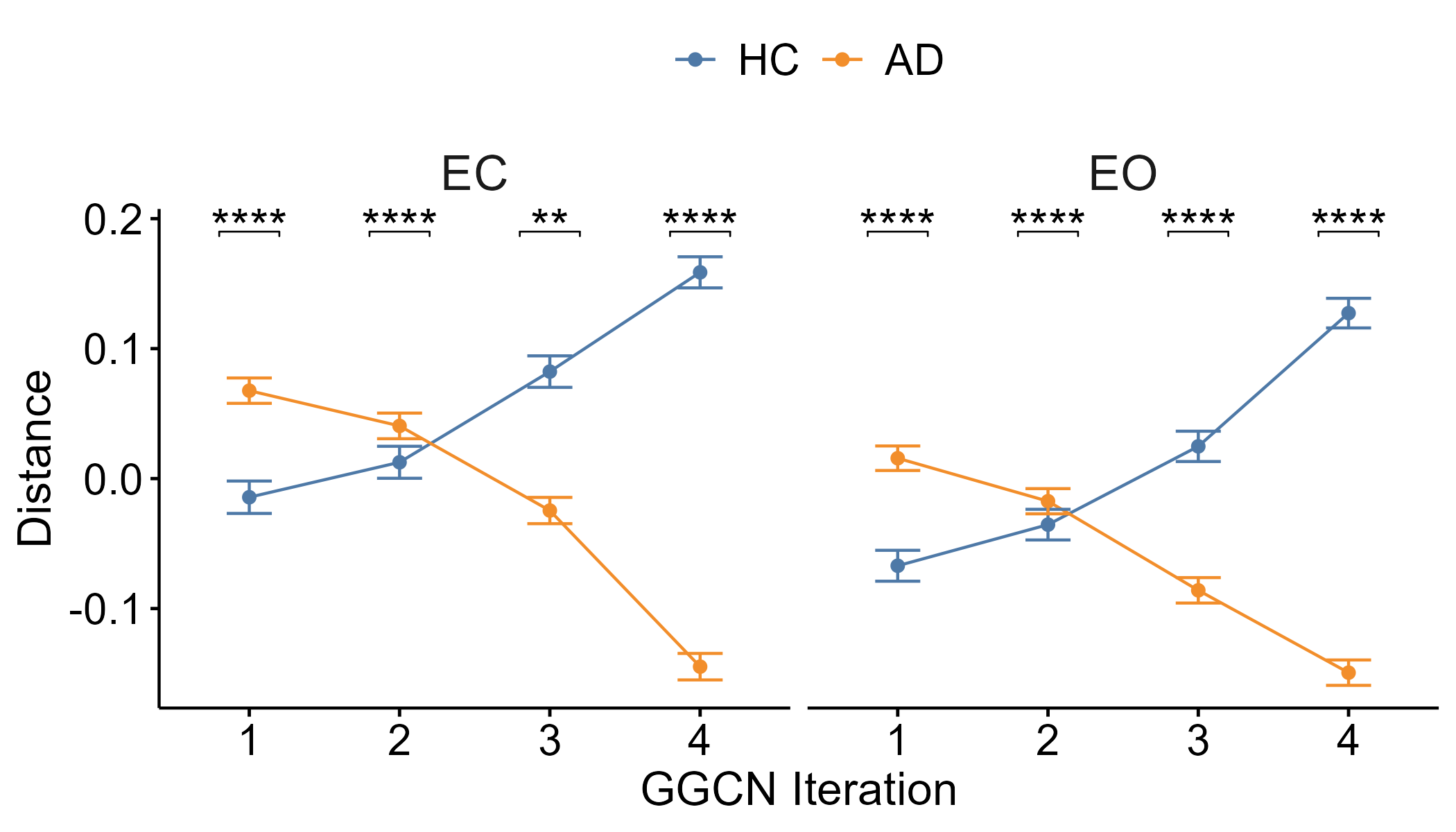}
    \caption{The average distance between initial node embedding and updated node embeddings shows the amount of information retained in each iteration of GGCN, i.e. going from local to global information. The asterisks denote the p-value of non-parametric Mann-Whitney U tests comparing the average distance between AD and HC cases and EC and EO conditions.}
    \label{fig:gated_information}
\end{figure}

Next, the role of the gating mechanism is elucidated by analysing the amount of information gathered at each scale, i.e. iteration of GGCN (Fig. \ref{fig:gated_information}). We measure this by computing the average Euclidean distance between the initial and updated node embedding at each iteration, i.e. $\mathbf{x}_i^{(r)}$ and $\mathbf{m}_i^{(r+1)}$ in Eq. \ref{eq: GGCN}. For instance, a small distance means a small amount of information was gathered at that scale.
Local information contributes highly to the node embeddings of the AD cases, and then the degree of contributions linearly decreases with increasing graph scale. The opposite pattern is observed for HC cases, where the later iterations influence the node embeddings. This highlights the degradation of global and distributed information caused by AD since the model can efficiently learn with fewer iterations, i.e. most information is obtained from the first three iterations.

\begin{figure}[ht]
    \centering
    \includegraphics[width = 1.05\linewidth]{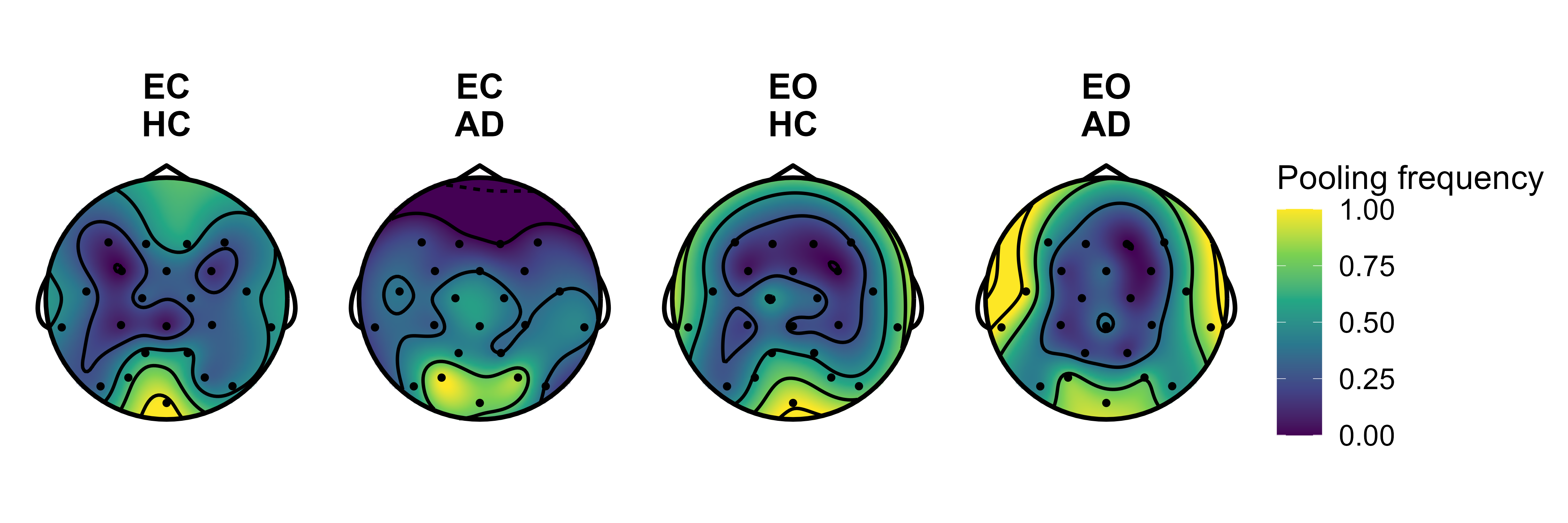}
    \caption{The average probability of a node being included in the coarsened graph by the ASAP node pooling module for AD and HC cases across EC and EO conditions. Averaged from all samples and min-max normalised.}
    \label{fig:cluster_probability}
\end{figure}

\begin{figure}[ht]
    \centering
    \includegraphics[width = 1.05\linewidth]{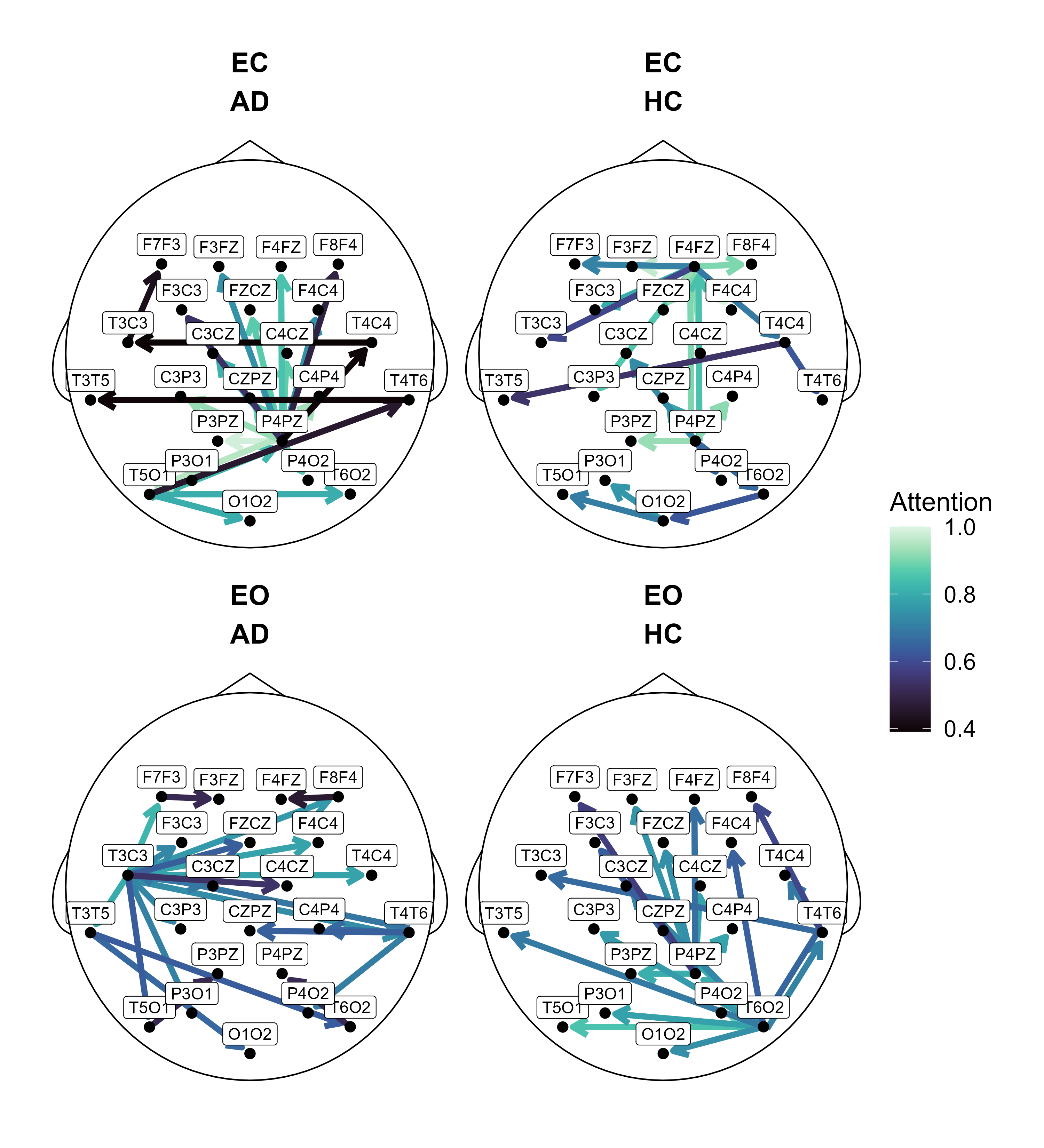}
    \caption{Attention scores learned by the node pooling module ($a_{ij}$ in Eq. \ref{eq: attention}), indicating the amount of information transferred from the source node into a cluster centred at the target node. Averaged for all AD and HC cases across EC and EO conditions (single strongest edge preserved for each target cluster node).}
    \label{fig:cluster_attention}
\end{figure}

\subsubsection{Node pooling module}
The node pooling mechanism can be exploited to derive two explanations. First, we analyse the frequency with which each node is included in the coarsened graph, i.e. pooling frequency (Fig. \ref{fig:cluster_probability}). Second, cluster attention scores (i.e. $a_{ij}$ in Eq. \ref{eq: attention}) can be used to identify important hubs that are highly represented in the clusters learned by the node pooling module (Fig. \ref{fig:cluster_attention}).

The nodes in parieto-occipital regions are consistently selected with high pooling frequency for AD and HC cases across both EC and EO conditions (Fig. \ref{fig:cluster_probability}). Additionally, in EC condition, HC cases frequently select frontal nodes while AD cases tend to select central nodes. In contrast, in the EO condition, there seems to be more variation in the pooling frequency, with temporal nodes having a high pooling frequency for AD and HC cases.

Note that the nodes of the pooled graphs are, in fact, cluster embeddings, i.e. attention weighted sum of node embeddings (Eq. \ref{eq: attention weighted sum}). We visualise the nodes with the highest attention scores of each cluster to highlight important hubs (Fig. \ref{fig:cluster_attention}). The attention scores are directed edges from a source node, transferring information to the cluster centred at the target node. Alternatively, these scores can be interpreted as a cluster membership strength. This information transfer should be interpreted as information flow within the model and most likely does not reflect an information flow within the brain.

In EC, AD cases show a large hub at the P4PZ node with strong long-distance and short-distance to various nodes. Additionally, there is a smaller hub at the T5O1. Similarly, in EO, AD cases have a large hub at the T3C3 node and a smaller one at the T4T6 node. In contrast, HC cases do not have any apparent hubs in the EC condition, with only a small hub at the P4PZ node. The attention links also seem to be rather short-distance. In the EO condition, HC cases show a large hub at the T6O2 node and smaller hubs at the P4PZ and T4T6 nodes.

This variance between EC and EO conditions displayed in the pooling frequency and attention scores suggests a plausible answer to why it is challenging for the model to learn joint representation in the EC+EO combined condition. We speculate this is caused by the additional dynamics introduced by the visual processing during the EO condition.

\subsubsection{Feature masking}
\begin{figure}[ht]
    \centering
    \includegraphics[width = 0.98\linewidth]{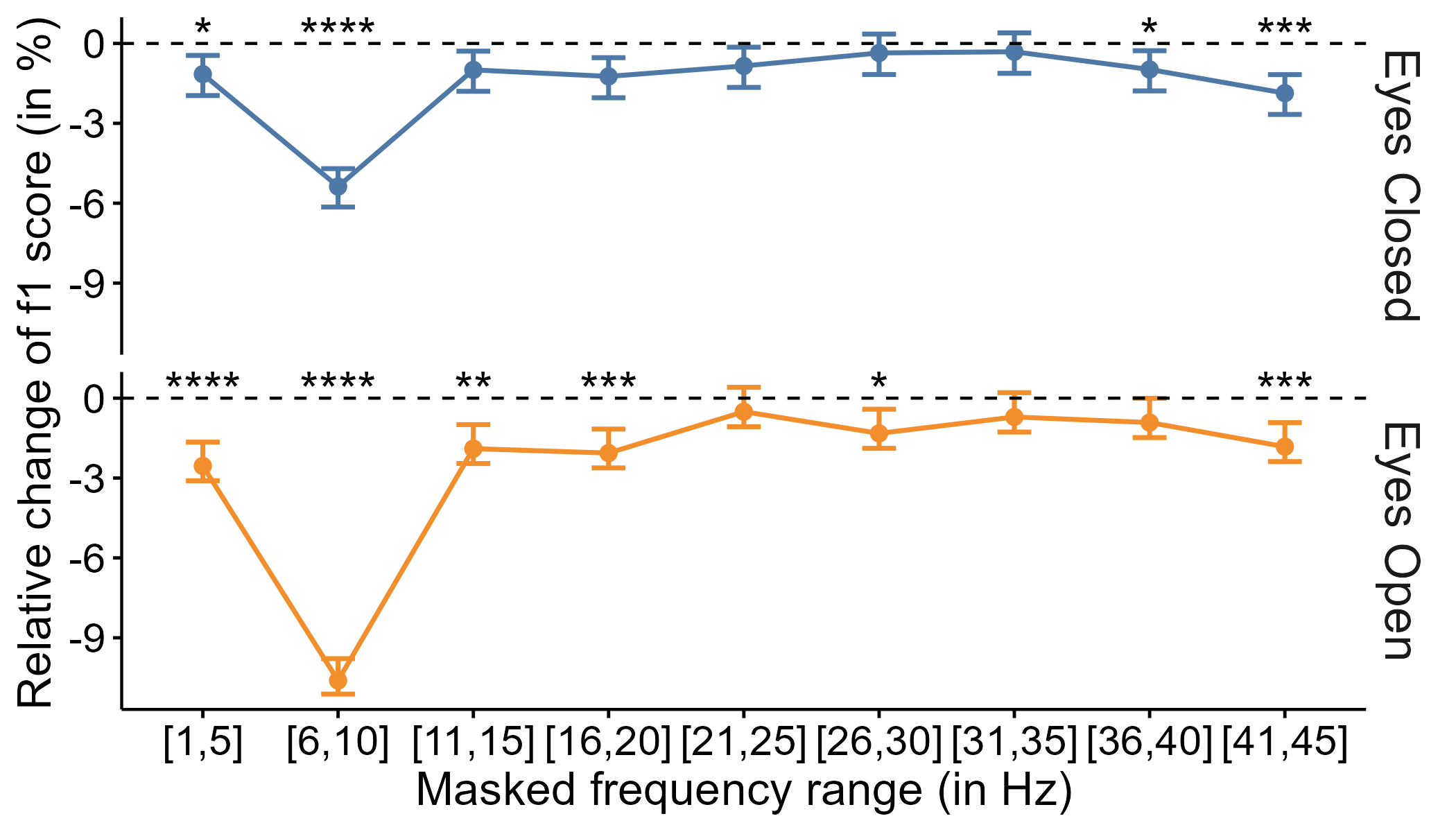}
    \caption{Relative change in F1 score when parts of node features are masked, showing the importance of frequency components for the classification task for eyes closed and eyes open conditions. The asterisks denote the p-value of non-parametric Mann-Whitney U tests comparing whether the relative change is significantly different from 0.}
    \label{fig:frequency_effect}
\end{figure}

We utilise feature masking to elucidate the importance of the frequency components summarised at each node by the node feature vector, i.e. PSD. In this, values at a selected part of the node feature vectors are replaced by zeroes and the model is retrained on this modified dataset. The relative reduction in f1 scores was then measured and visualised in Fig \ref{fig:frequency_effect} for EC and EO conditions.

In both EC and EO conditions, the frequencies between 6 and 10 Hz are the most important since their masking reduced performance by 4.82\% and 9.18\%, respectively. This fits well with the well-described increase of power as well as functional connectivity in AD within these frequencies corresponding to $\theta$ and low $\alpha$ bands \cite{babiloni2016synchronisationreview, klepl2022bispectrum}. Similarly, masking of the $[1,5]$, $[36,40]$ and $[41,45]$ frequency ranges results in a significant performance decrease in both EC and EO. Additionally, in EO condition, the $[11,15]$, $[16,20]$ and $[26,30]$ frequency ranges produce a significant performance decrease.

\subsection{Limitations and future work}
Although our approach achieves competitive performance, we identify a few drawbacks. First, the relatively small size of our dataset imposes a limit on fitting complex models. We address this issue by segmenting the EEG signals into short windows. The short window length means that the model might not be able to represent information from low-frequency components of the signal. 

Next, we do not explore alternative node feature representations beyond PSD in this study. PSD is merely a linear frequency-domain representation of the signal. Including time-domain and nonlinear information in the node features might improve the expressiveness of the model. Similarly, the proposed graph learning mechanism is limited to linear coupling patterns because (1) it is inferred from the node features and (2) it is expressed as Pearson's correlation coefficient. Future work should explore other forms of FC that might be integrated into the graph learning mechanism and study ways to include more complex frequency-dependent coupling information.

Finally, the model architecture might be limited by the relatively large number of hyper-parameters that need to be optimised. However, this limitation should be mitigated by utilising a validation set during the optimisation. Moreover, we explore the model stability with respect to some of the important hyperparameters in the parameter sensitivity experiments. These suggest that the achieved performance of the proposed model is not limited purely to the optimal values of the hyperparameters.

\section{Conclusion}
This work proposes a novel graph learning model that performs highly in the AD diagnosis task. Additionally, we show that the model produces robust and clinically relevant explanations for its predictions via the novel graph structure learning module and the node pooling mechanism. Finally, we highlight the importance of utilising the gating mechanism within a message-passing encoder. This allows the model to accurately represent the multiscale distributed network disruptions displayed in the AD cases.

\section*{Acknowledgement}
The EEG data was funded by a grant from Alzheimer’s Research UK (ARUK-PPG20114B-25). The views expressed are those of the author(s) and not necessarily those of the NHS, the NIHR or the Department of Health. The work was supported by A*STAR, AI, Analytics and Informatics (AI3) Horizontal Technology Programme Office (HTPO) seed grant C211118015. 

\newpage
\bibliographystyle{unsrt}
\bibliography{ref}

\newpage
\appendix
\include{supplements_preprint}

\end{document}

%% file: supplements_preprint.tex
\section{Hyperparameters of proposed model}
The optimised and allowed values of the various hyperparameters of the proposed adaptive gated graph convolutional network (AGGCN) are reported in Tables \ref{table:hyperparameters} and \ref{table:hyperparameters_range}, respectively.

\begin{table}[!ht]
\centering
\caption{Hyper-parameter values of the optimised model}
\scalebox{0.8}{
\begin{tabular}{cccccc}
  \hline
$L_{CNN}$ & kernel size & CNN filters & $h_{CNN}$ & $drop_{CNN}$ & $k_{KNN}$ \\ 
  \hline
  \hline
  1 & 4 & 84 & 403 & 0.024 & 16 \\ 
  \hline
  $R$ & $h_{GNN}$ & activation & aggregation & $drop_{GNN}$ &  \\ 
  \hline
  \hline
  4 & 372 & Tanh & mean & 0.9 &  \\ 
  \hline
  $k_{pool}$ & $drop_{pool}$ & negative slope & $L_{MLP}$ & $h_{MLP}$ & $drop_{MLP}$ \\ 
  \hline
  \hline
  3 & 0.75 & 0.085 & 3 & 16 & 0 \\ 
  \hline
  learning rate & momentum & weight decay & $\gamma$ & $\sigma$ & $p_{noise}$ \\
  \hline
  \hline
  0.063 & 0.859 & 0.076 & 0.896 & 0.346 & 0.1 \\ 
   \hline
\end{tabular}
}
\label{table:hyperparameters}
\end{table}

\begin{table}[!ht]
\centering
\caption{Hyper-parameter value ranges allowed during optimisation} 
\begin{tabular}{cc}
  \hline
Hyperparameter & Values \\ 
  \hline
$L_{CNN}$ & $[1,\dots,4]$ \\ 
  kernel size & $[2,\dots,4]$ \\ 
  CNN filters & $[16,\dots,100]$ \\ 
  $h_{CNN}$ & $[16,\dots,1024]$ \\ 
  $drop_{CNN}$ & $[0, 0.9]$ \\ 
  $k_{KNN}$ & $[1,\dots,23]$ \\ 
  $R$ & $[1,\dots,10]$ \\ 
  $h_{GNN}$ & $[16,\dots,1024]$ \\ 
  activation & $ReLU, Tanh, ELU, LeakyReLU$ \\ 
  aggregation & $add, mean, max$ \\ 
  $drop_{GNN}$ & $[0, 0.9]$ \\ 
  $k_{pool}$ & $[1,\dots,23]$ \\ 
  $drop_{pool}$ & $[0, 0.9]$ \\ 
  negative slope & $[0, 0.5]$ \\ 
  $L_{MLP}$ & $[1,\dots,5]$ \\ 
  $h_{MLP}$ & $[16,\dots,2048]$ \\ 
  $drop_{MLP}$ & $[0, 0.9]$ \\ 
  learning rate & $[0.001, 0.1]$ \\ 
  momentum & $[0, 0.9]$ \\ 
  weight decay & $[0, 0.1]$ \\ 
  $\gamma$ & $[0.8, 0.95]$ \\ 
  $\sigma$ & $[0, 0.5]$ \\ 
  $p_{noise}$ & $[0, 0.6]$ \\ 
   \hline
\end{tabular}
\label{table:hyperparameters_range}
\end{table}

\section{Parameter sensitivity experiments}
\begin{figure}[!ht]
    \centering
    \includegraphics[width = 0.98\linewidth]{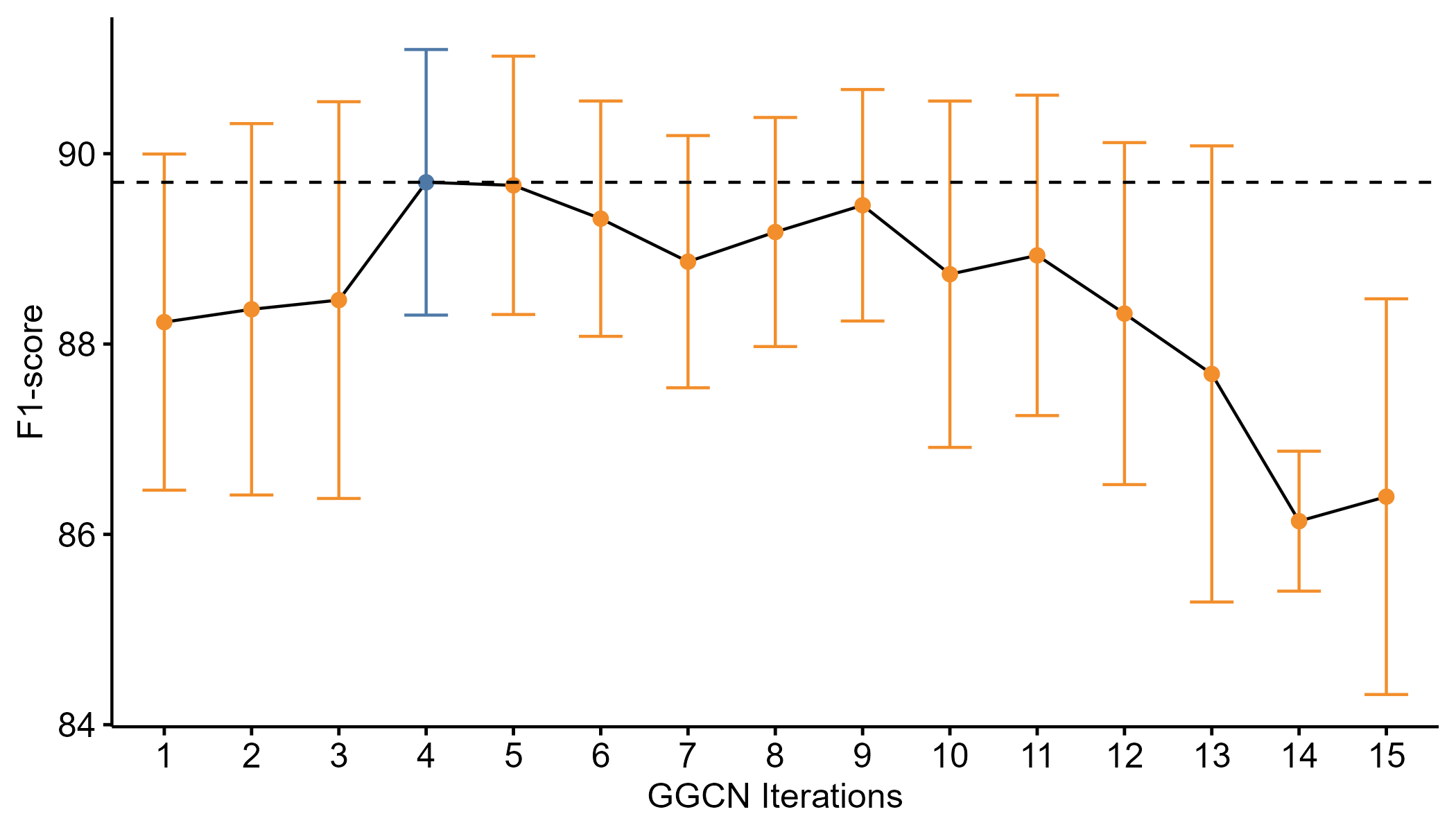}
    \caption{Sensitivity of the proposed model to the number of iterations of the GGCN encoder. The error bars show the standard deviation of accuracies measured across 10 repetitions. The optimal value showed in blue.}
    \label{fig:GGCN_iterations}
\end{figure}

\begin{figure}[!ht]
    \centering
    \includegraphics[width = 0.98\linewidth]{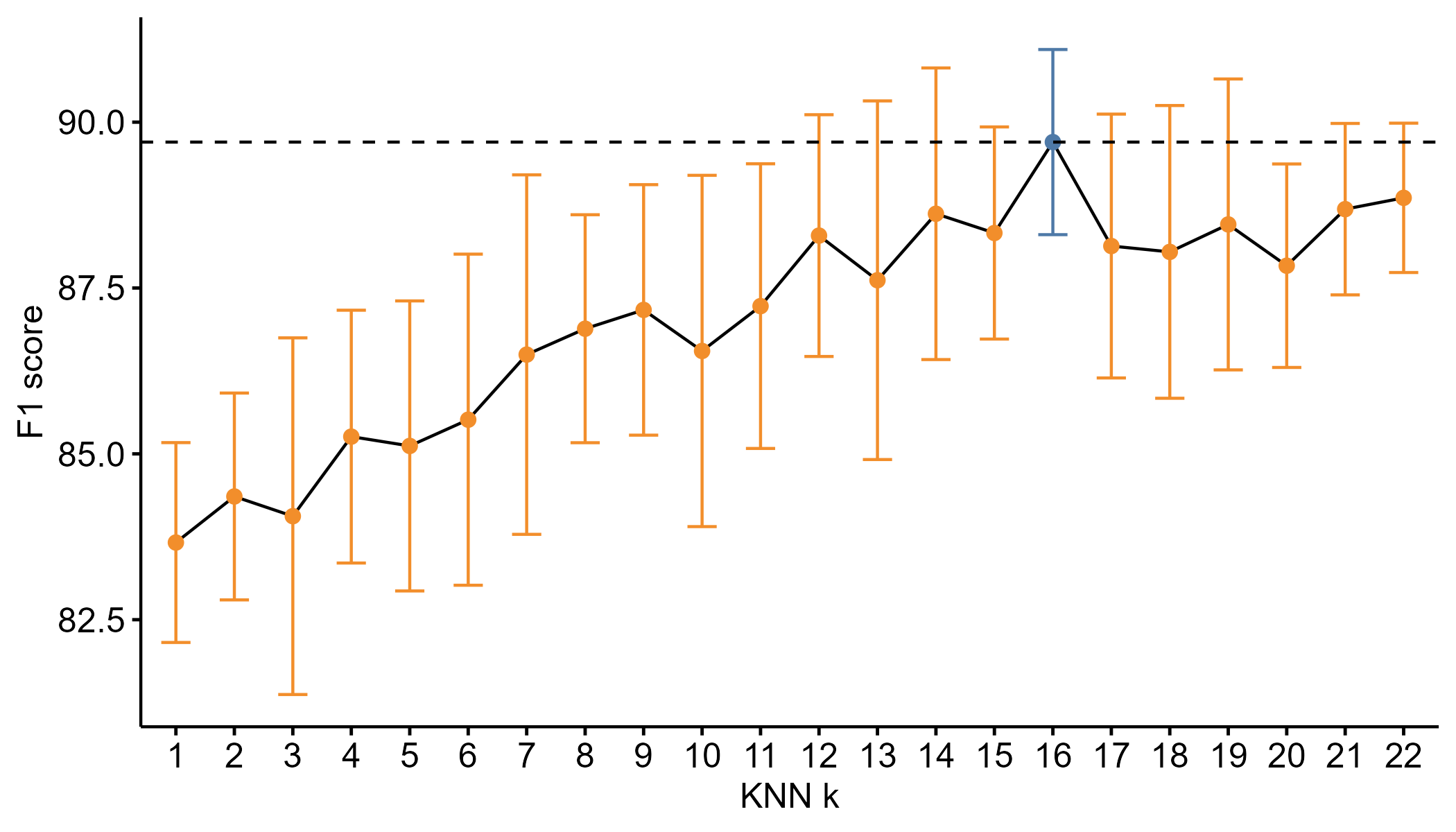}
    \caption{Sensitivity of the proposed model to the k-nearest-neighbour edges kept in the learned graph structure. The error bars show the standard deviation of accuracies measured across 10 repetitions. The optimal value showed in blue.}
    \label{fig:KNN}
\end{figure}

\begin{figure}[!ht]
    \centering
    \includegraphics[width = 0.98\linewidth]{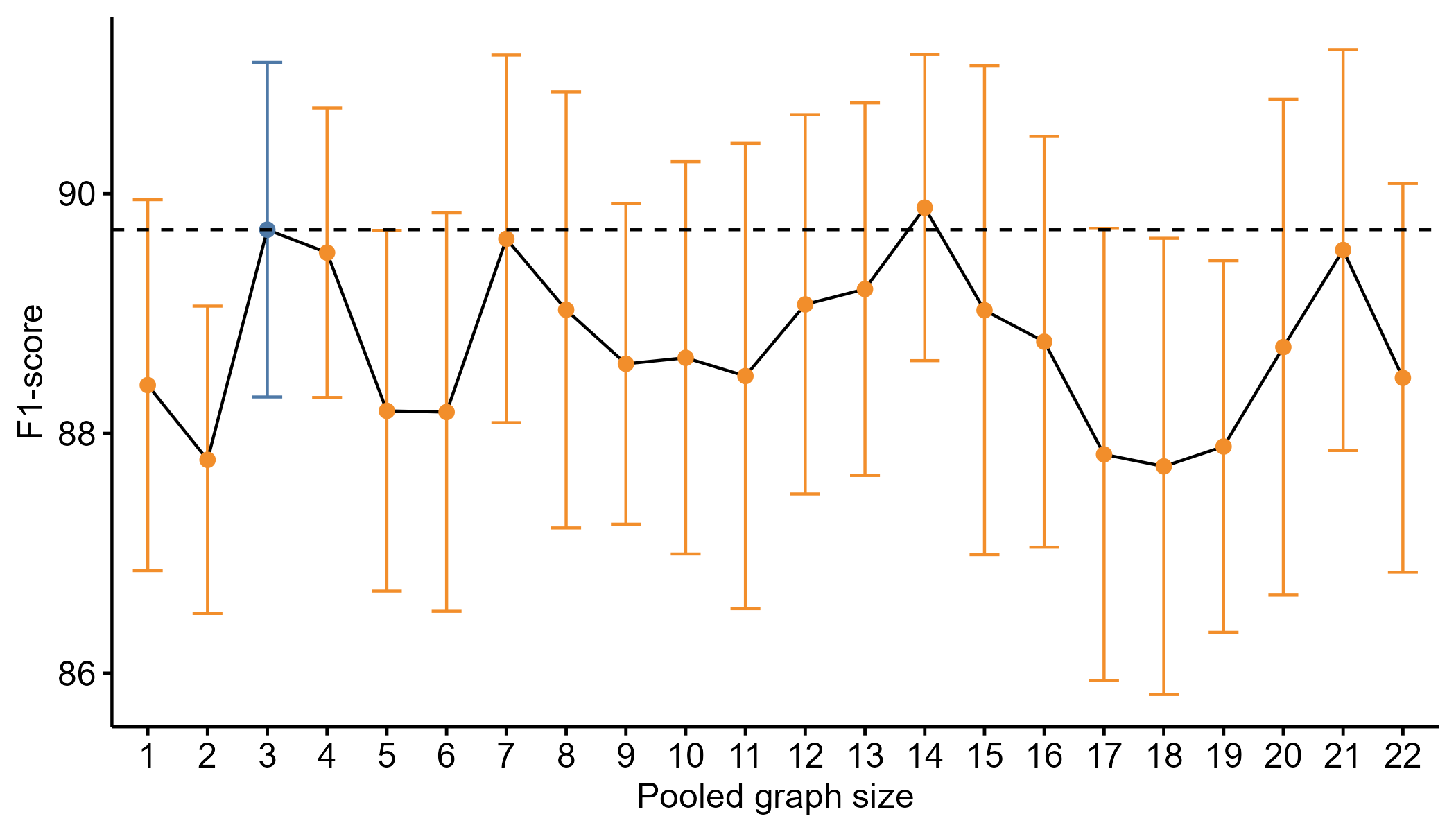}
    \caption{Sensitivity of the proposed model to the size of the pooled graph. The error bars show the standard deviation of accuracies measured across 10 repetitions. The optimal value showed in blue.}
    \label{fig:pooled_size}
\end{figure}

\begin{figure}[!ht]
    \centering
    \includegraphics[width = 0.98\linewidth]{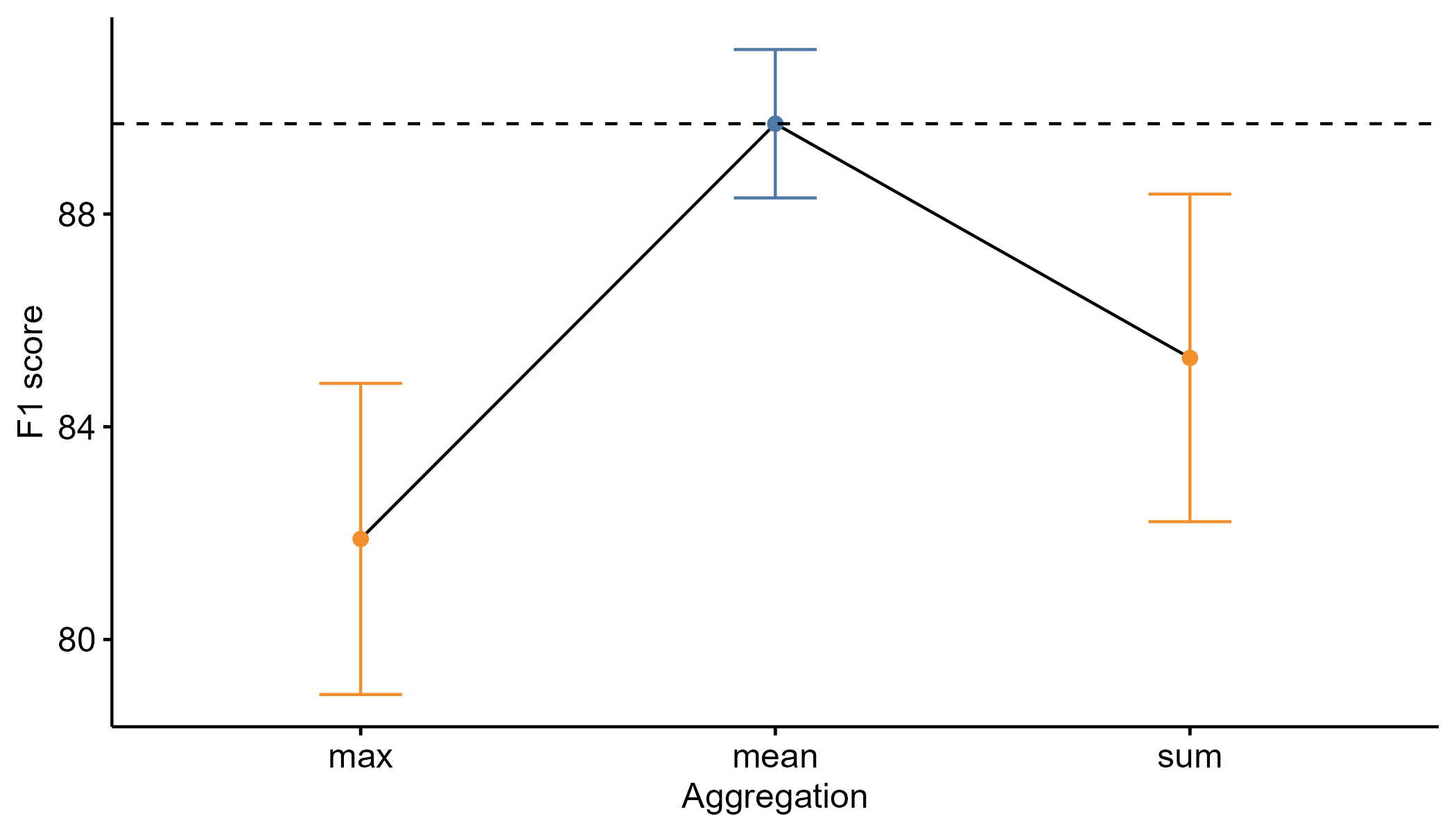}
    \caption{Sensitivity of the proposed model to the choice of the aggregation function. The error bars show the standard deviation of accuracies measured across 10 repetitions. The optimal value showed in blue.}
    \label{fig:aggregation}
\end{figure}

Multiple parameter sensitivity experiments were performed to test the influence of the selected crucial hyperparameters of AGGCN. The results of these experiments are reported in Figures \ref{fig:GGCN_iterations}, \ref{fig:KNN}, \ref{fig:pooled_size} and \ref{fig:aggregation} for the number of GGCN iterations, K-nearest neighbour edges kept in the sparse learned graph structure, size of the coarsened (pooled) graph and aggregation function, respectively.

\section{Explainability of AGGCN: Adjacency-based visualisations}
\begin{figure}[!ht]
    \centering
    \includegraphics[width = 0.98\linewidth]{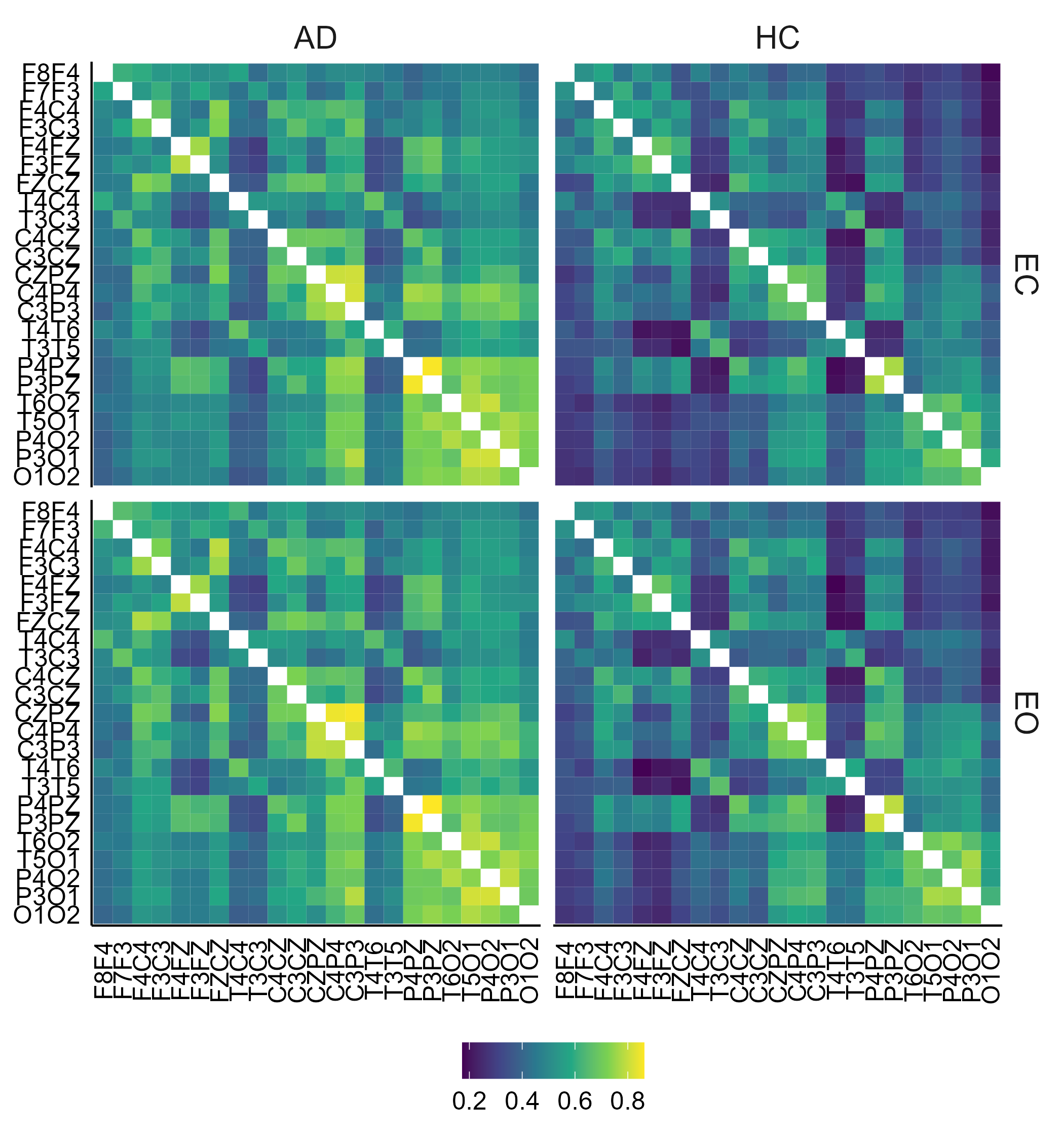}
    \caption{Average adjacency matrix of learned graphs of AD and HC cases in EC and EO conditions.}
    \label{fig:learned_adjacency}
\end{figure}

\begin{figure}[!ht]
    \centering
    \includegraphics[width = 0.98\linewidth]{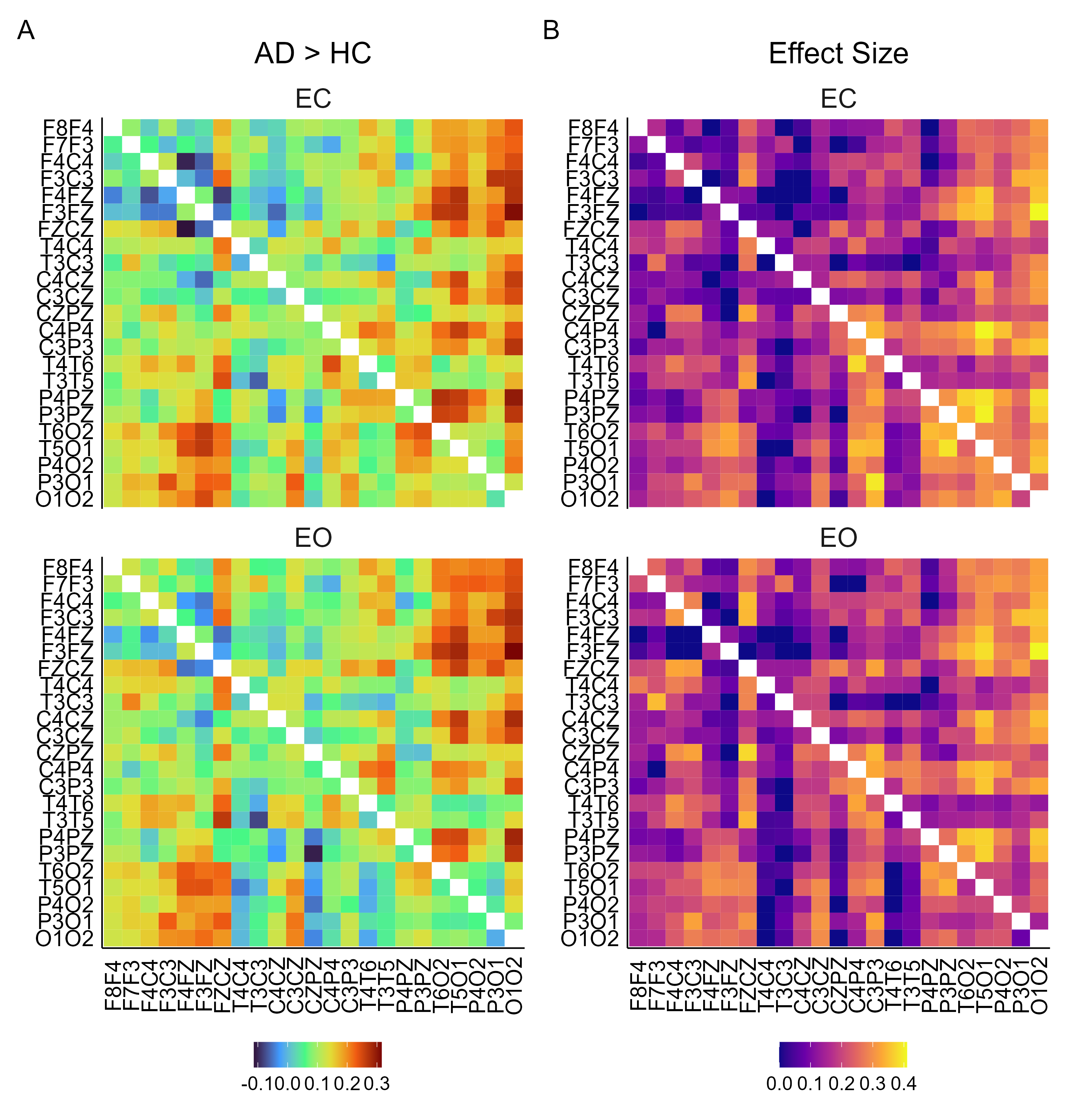}
    \caption{Difference of averaged adjacency matrices of learned graphs of AD and HC cases ($AD-HC$) in EC and EO conditions (A).} (B) The effect size for the non-parametric Mann-Whitney U tests comparing AD and HC with values set to 0 where $\text{p-value} > 0.05$,
    \label{fig:learned_adjacency_difference}
\end{figure}

\begin{figure}[!ht]
    \centering
    \includegraphics[width = 0.98\linewidth]{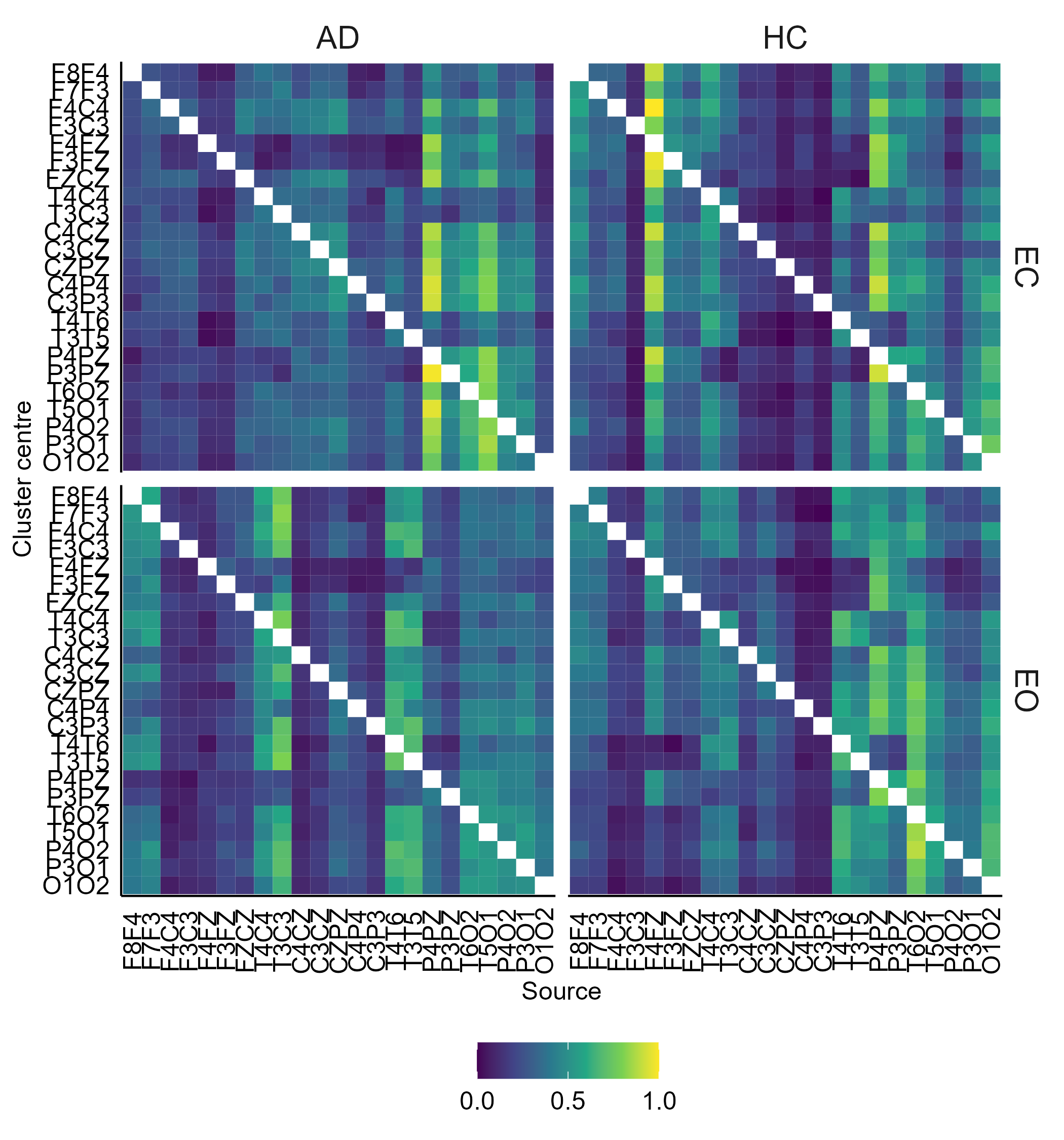}
    \caption{Average adjacency matrix of attention scores obtained by the node pooling module for AD and HC cases across EC and EO conditions.}
    \label{fig:attention_adjacency}
\end{figure}

The main manuscript shows the AGGCN-learned graphs and the node pooling patterns as a graph. In order to facilitate a different view of the same results, we report the averaged adjacency matrices in Figure \ref{fig:learned_adjacency} that correspond to Figure 3 in the main text. Similarly, we report the differences between the learned graphs together with effect sizes (Wilcox permutation effect size) to quantify the strength of these differences (Figure \ref{fig:learned_adjacency_difference}, corresponding to Figure 4 in the main text). Finally, we report an adjacency-like view of the node pooling attention scores (Figure \ref{fig:attention_adjacency}) corresponding to Figure 8 in the main text.

